\documentclass[12pt]{elsarticle}

\usepackage{lineno,hyperref}
\usepackage[version=4]{mhchem}
\usepackage{siunitx}
\modulolinenumbers[5]
\graphicspath{ {./pics/} }
\usepackage{textcomp}
\usepackage[flushleft]{threeparttable}
\usepackage{makecell}
\usepackage{subfigure}
\usepackage{graphicx}
\usepackage{lipsum}
\usepackage{textcomp}
\usepackage{nomencl} % Nomenclature package
\makenomenclature

\begin{document}

\title{Battery Cloud with Advanced Algorithms}
% \tnotetext[mytitlenote]{Fully documented templates are available in the elsarticle package on \href{http://www.ctan.org/tex-archive/macros/latex/contrib/elsarticle}{CTAN}.}

%% Group authors per affiliation:
% \chapterauthor{Xiaojun Li* \\ David Jauernig \\ Mengzhu Gao \\ Trevor Jones}
% \chapteraff{
% Gotion Inc\\
% 48660 Kato Road, Fremont, California, USA\\
% \href{mailto:t.li@gotion.com}{\textit{t.li@gotion.com}}}

% \fnref{myfootnote}

%% or include affiliations in footnotes:
\author{Xiaojun Li\corref{mycorrespondingauthor}}
\author{David Jauernig}
\author{Mengzhu Gao}
\author{Trevor Jones}

% \cortext[mycorrespondingauthor]{Corresponding author}
% \ead{support@elsevier.com}

\cortext[mycorrespondingauthor]{Corresponding author (\href{mailto:t.li@gotion.com}{t.li@gotion.com})}

\address{
Gotion Inc, 48660 Kato Road, Fremont, California, USA \\
\{t.li,d.jauernig,m.gao,t.jones\}@gotion.com}

% \affil[1]{organization={Gotion Inc},%Department and Organization
%             addressline={48660 Kato Road}, 
%             city={Fremont},
%             postcode={94043}, 
%             state={CA},
%             country={USA}}
% \affil[2]{{t.li,d.jauernig,m.gao,t.jones}@gotion.com}
            
% \begin{abstract}

% \end{abstract}

% \begin{keyword}
% \texttt{elsarticle.cls}\sep \LaTeX\sep Elsevier \sep template
% \MSC[2010] 00-01\sep  99-00
% \end{keyword}

\begin{abstract}
Energy storage battery plays a key role in modern interconnected energy networks. Recent development of Internet of Things (IoT) has enabled traditional battery management system to evolve into Battery Cloud. A Battery Cloud or cloud battery management system leverages the cloud computational power and data storage to improve battery safety, performance, and economy. This work will present the Battery Cloud that collects measured battery data from electric vehicles and energy storage systems. Advanced algorithms are applied to improve battery performance. Using remote vehicle data, we train and validate an artificial neural network to estimate pack SOC during vehicle charging. The strategy is then tested on vehicles.
Furthermore, high accuracy and onboard battery state of health estimation methods for electric vehicles are developed based on the differential voltage (DVA) and incremental capacity analysis (ICA). Using cycling data from battery cells at various temperatures, we extract the charging cycles and calculate the DVA and ICA curves, from which multiple features are extracted, analyzed, and eventually used to estimate the battery’s state of health. For battery safety, a data-driven thermal anomaly detection method is developed. The method can detect unforeseen anomalies such as thermal runaways at the very early stage. Potential applications of battery cloud also include areas such as battery manufacture, recycling, and electric vehicle battery swap.
% ^ <tonylee2016@gmail.com> 2018-05-27T00:14:44.455Z.
\end{abstract}


\maketitle

\nomenclature{SOC}{State of Charge}
\nomenclature{C-rates}{ the rate of discharge/charge based on the capacity of the battery} 
\nomenclature{SOP}{State of Power} 
\nomenclature{SOH}{State of Health} 
\nomenclature{SOH}{State of Energy} 
\nomenclature{SOX}{State of Charge/Energy/Power/Health, battery core algorithm} 
\nomenclature{BMS}{Battery Management System} 
\nomenclature{IC}{Integrated Circuitry} 
\nomenclature{\textmu{C}}{Micro-controller} 
\nomenclature{ANN}{Artificial Neural Networks} 
\nomenclature{EV}{Electric Vehicle} 
\nomenclature{ESS}{Energy Storage System} 
\nomenclature{IoT}{Internet of Things} 
\nomenclature{DVA}{Differential Voltage Analysis} 
\nomenclature{ICA}{Incremental Capacity Analysis} 
\nomenclature{HDFS}{Hadoop File System} 
\nomenclature{TSDB}{Timeseries Database} 
\nomenclature{ML}{Machine Learning} 
\nomenclature{ECM}{Equivalent Circuitry Model} 
\nomenclature{KF}{Kalman Filter} 
\nomenclature{HIL}{Hardware in the Loop}
\nomenclature{SEI}{Solid Electrolyte Interphase}
\nomenclature{NMC}{Nickel Manganese Cobalt} 
\nomenclature{LLI}{Loss of Lithium Inventory} 
\nomenclature{LAM\textsubscript{A}}{Loss of active material of the anode} 
\nomenclature{LAM\textsubscript{C}}{Loss of active material of the cathode} 
\nomenclature{EOL}{End of Life} 
\nomenclature{LUT}{Look-Up-Table} 
\nomenclature{TR}{Thermal Runaway} 

\printnomenclature

\tableofcontents

\linenumbers

% \setcounter{chapter}{5}
% \chapter{Battery Cloud with Advanced Algorithms}

\section{Introduction}

\begin{figure}[t]
\includegraphics[width=\textwidth]{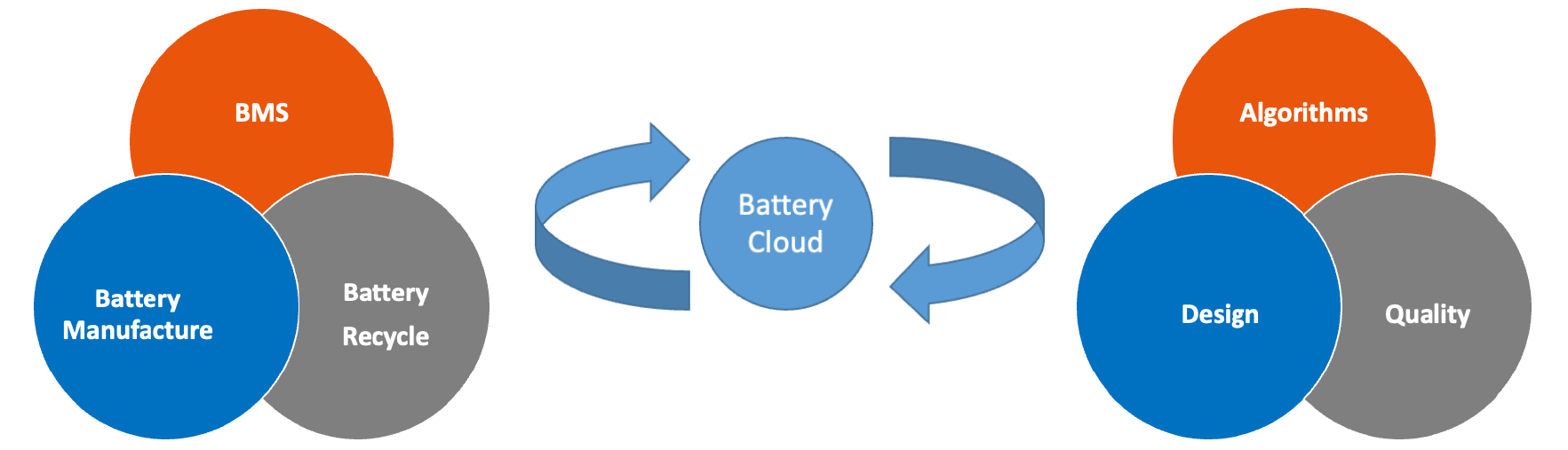}
\caption{Battery Cloud is at the center of the success of the future battery industry.}
\end{figure}

Batteries play an essential role in the rapid development of transportation electrification and energy storage systems\cite{Dunn2011}. Lithium-ion batteries are known for their high energy/power density and low self-discharge. They are becoming more available as the manufacturing cost continues to improve. Large-scale energy storage systems consist of MWh/GWh batteries that continuously operate under different weather conditions. Electric vehicle batteries are subject to road harshness, different driving behavior, and frequent high current fast charges. These applications call for batteries to become more reliable, safe, and predictable. As such, monitoring and control of Li-ion batteries become more critical. Battery algorithms, such as SOC and SOH, deliver important information about battery charge and health. This information is critical for maintaining optimal operations of modern energy networks. For example, inaccurate estimation of SOC will force the battery system to reduce charge/discharge power or completely shut off, which subsequently affects the grid stability. Based on accurate SOH estimation, a modern energy network can reduce the risk of battery failure. Early thermal anomaly detection can foresee thermal runaway, which is catastrophic for energy networks.   
 
 \setlength{\tabcolsep}{20pt}
\renewcommand{\arraystretch}{1.5}

 \begin{table}[b]
\centering 
  \begin{threeparttable}
    \caption{Lithium-ion Battery for Energy Storage Applications}\label{battery_types}
    \begin{tabular}{ccccc}
    \hline
    Types  & Specific Energy & Life Cycle & Cost\\
    \hline
    NMC       & over 150 Wh/kg   & 1200 &  \$ 403/kWh\\
    LFP     &  over 110 Wh/kg   & 2000 &  \$ 274/kWh\\
    \hline  % Please only put a hline at the end of the table
    \end{tabular}
  \end{threeparttable}
    \begin{tablenotes}
      \small
      \item Note: cost estimation based on 10MW, without warranty, insurance, maintenance.\cite{PacificNorthwestNationalLaboratory2022Lithium-ionNMC}
    \end{tablenotes}
\end{table}

As of now, conventional onboard battery management systems (BMS) are used for monitoring and control. A BMS includes embedded micro-controllers (\textmu{C}) and peripheral integrated circuitry (IC). Usually, the BMS collects voltage, current, and temperature measurement with dedicated sensing ICs that communicates with a main \textmu{C}, which process the measurements and perform various functions, such as SOX estimation, and diagnostics, protection, control, and thermal management. Nevertheless, the micro-controllers are designed to handle simple tasks and have the minimal computing power and memory size. It prevents the onboard BMS from executing advanced algorithms. For example, artificial neural networks (ANN) are frequently used for SOC estimation\cite{Lombardo2021ArtificialReality}. As we will show later in this chapter, an onboard BMS might run a trained neural network. However, the ANN must be carefully designed to reduce CPU and RAM impact. Although the BMS receives numerous data from the measurements of hundreds of cells that it monitors, these data are not stored due to the lack of onboard data storage, making incremental learning impossible for the onboard BMS.

With the further development of IoT\cite{Xu2014InternetSurvey}, future BMS is expected to be cloud-connected (Battery Cloud). As a result, battery data can be seamlessly uploaded and stored in a cloud data platform\cite{Voltaiq,HowAWS}, and the power of cloud computing resources can be leveraged. The cloud computational power and data storage can support advanced algorithms, such as machine learning algorithms improve battery safety, performance, and economy. There are several significant advantages. Firstly, the cloud database has battery data from not just one pack but numerous EV/ESS battery packs, allowing a massive amount of data to be used for extensive data analysis and machine learning. Secondly, cloud computing allows complicated algorithms to be executed in real-time, which is not possible for onboard \textmu{C}. Thirdly, the cloud platform allows data collection and feedback from batteries throughout the entire life cycle. The other battery processes and applications also benefit from the battery cloud, such as manufacturing, second life usage, and recycling. 

This chapter aims to provide an overview of Battery Cloud, including the essential infrastructure and software components. We also discuss important topics for batteries, including the underlying causing mechanism, effects, and corresponding algorithms. The remaining chapter is organized as follows. The first section discusses the critical components of a Battery Cloud. Then, in the following sections, we overview the critical areas regarding battery performance, health, and safety: State-of-Charge estimation, State-of-Health estimation, and thermal runaway/anomaly detection. We also present corresponding algorithms that were developed with the Battery Cloud. In the first section, we train and validate an artificial neural network (ANN) to estimate pack SOC during vehicle charging using remote vehicle data. The ANN is then implemented and tested by onboard BMS. It gives high accurate (<3\%) real-life vehicle testing results. In the second section, high accuracy (<\%5) and onboard battery state of health estimation (SOH) methods for electric vehicles are developed based on the differential voltage (DVA) and incremental capacity analysis (ICA). We extract the charging cycles and calculate the DVA and ICA curves using cloud data. Multiple features are extracted and analyzed to estimate the SOH. A data-driven thermal anomaly detection method is developed for battery safety in the last section. The method can detect unforeseen thermal anomalies at an early stage, more than 1hr ahead of the event.

\section{Battery in the Cloud}

\begin{figure}[t]
\includegraphics[width=\textwidth]{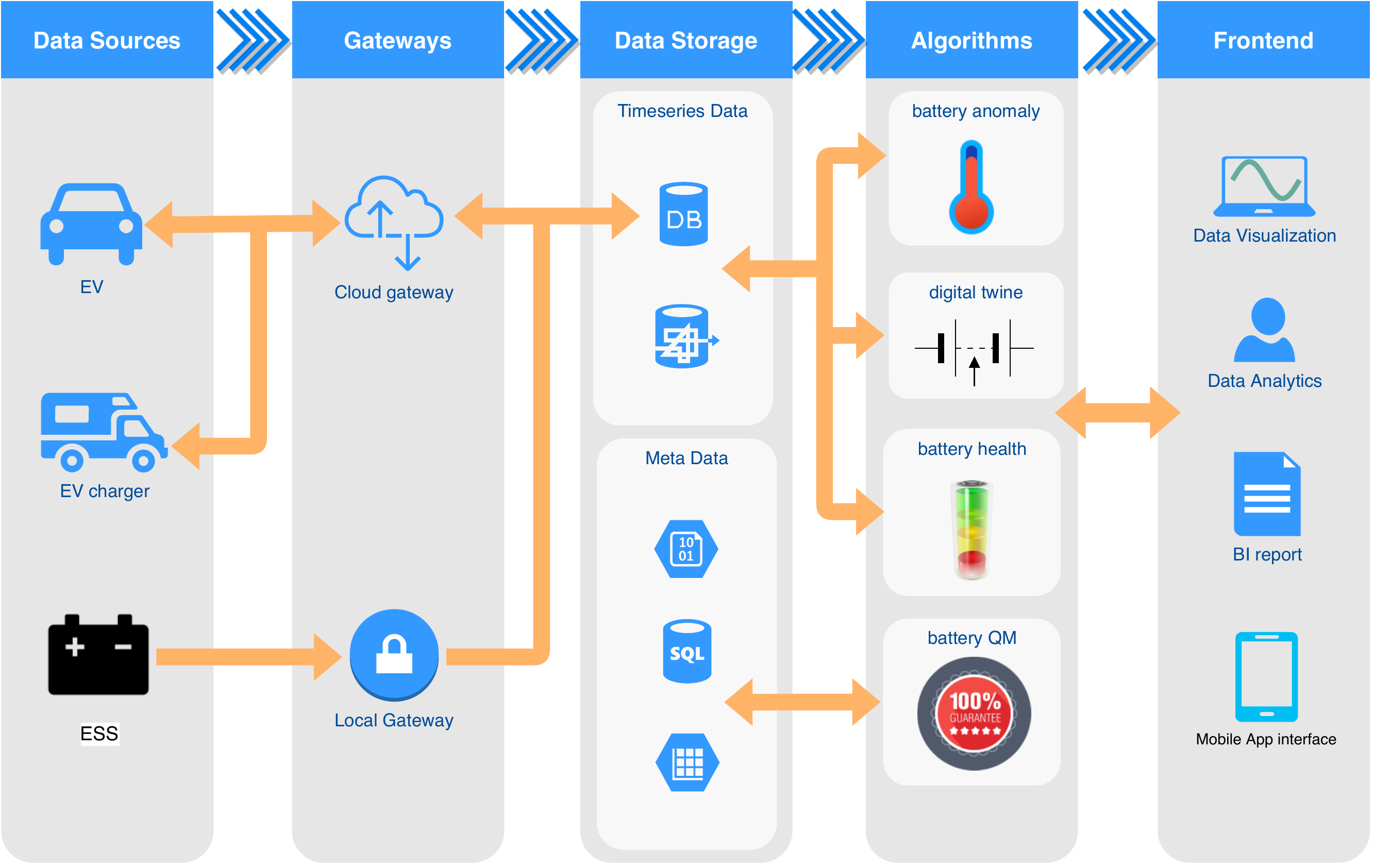}
\caption{Key hardware and software components and data flow of the Battery Cloud: data storage includes but not limit to (from top to bottom):  noSQL database, data storage, binary files, SQL database, spreadsheets.}
\end{figure}

\begin{table}[t]
\centering 
  \begin{threeparttable}
    \caption{Battery Data Sources}\label{datasrc}
    \begin{tabular}{lll}
    \hline
    Item  & Stages & Type of data \\
    \hline
    Cell  & \makecell[tl]{manufacture \\ testing } & \makecell[tl]{manufacture metadata \\ battery timeseries data}\\
    Pack & \makecell[tl]{assembly \\ testing } & \makecell[tl]{assembly metadata \\ battery timeseries data }\\
    EV\&{ESS} & \makecell[tl]{operating\\\\\\service} & \makecell[tl]{battery timeseries data \\ vehicle/grid timeseries data \\ vehicle/grid metadata \\ service record} \\
    Charger & \makecell[tl]{operating} & \makecell[tl]{charger timeseries data}\\
    Pack & \makecell[tl]{recycle} & \makecell[tl]{recyle metadata}\\
    \hline  % Please only put a hline at the end of the table
    \end{tabular}
  \end{threeparttable}
\end{table}

This section covers essential components of a Battery Cloud, including the database, data visualization, and algorithm/analytics.

\subsection{Data Sources and Connections}
Data are collected during different stages of the battery's life cycle, ranging from cell manufacture and module/pack assembly to vehicle driving/charging and pack recycling. There are numerous procedures for cell manufacture alone, including electrode mixing, coating, laser cutting, stack, and so many others\cite{Schnell2019DataProduction}, during which a significant amount of data is generated. Table ~\ref{datasrc} summarizes battery-related data based on the different devices and scenarios. The EV battery pack is equipped with a BMS, a wireless IoT component that transmits the collected data to the cloud via the 4G/5G network. These data will be collected via the internet, online or private gateways for charging stations. Because ESS power plants affect grid stability, they are subject to more stringent cybersecurity regulations. As a result, usually, ESSs are connected through a one-way, local gateway to ensure maximum security. Similarly, battery data from cell/pack testing equipment are uploaded via a secured, one-way gateway. However, the equipment may be controlled securely via the company's intranet.

\subsection{Database} 

\paragraph{Choosing the right database} For production big data platforms, Hadoop\cite{ApacheHadoop} is the prevailing choice. Hadoop is based on HDFS (Hadoop files system) and MapReduce (the programming model) that ensure good scalability, robustness, and high availability, all of which are essential requirements for a battery database. Besides, Hadoop has a complete ecosystem, including software stacks like Spark, Hbase, Kafka, Hive, and many others, making it easier to use and expand functionalities. There are also dedicated timeseries databases (TSDB), such as Influxdb, Timescale, and Prometheus. TSDB has built-in features for timeseries data, such as time-domain queries (integration, differential), retention policy, and others. This makes TSDB ideal for a small R\&D battery database. As TSDBs are being developed and improved actively, they will become more competitive against traditional databases in the future.

\paragraph{Database deployment}The database can be hosted on primes or on the cloud. Although on-primes deployment will theoretically give better control and security, it is often more expensive to maintain and scale. For cloud deployment, there are several models to consider. Iaas (Infrastructure as a Service) let the cloud provider handle hardware resources, where the company has complete control of software stacks. Popular IaaS providers are Amazon Web Services (AWS)\cite{AmazonService}, Microsoft Azure\cite{MicrosoftAzure}, and Google Cloud Platform (GCP)\cite{GooglePlatform}. In the PaaS (Data Platform as a Service) model, such as AWS EMR, the cloud provider also hosts basic software stacks, except for application software. The provider manages all software stacks in the SaaS (Software as a Service) model, such as Cloudera\cite{Cloudera} and Influxdata Cloud\cite{Influxdata}.

\subsection{Data Visualization} 

\begin{figure}[ht]
\includegraphics[width=\textwidth]{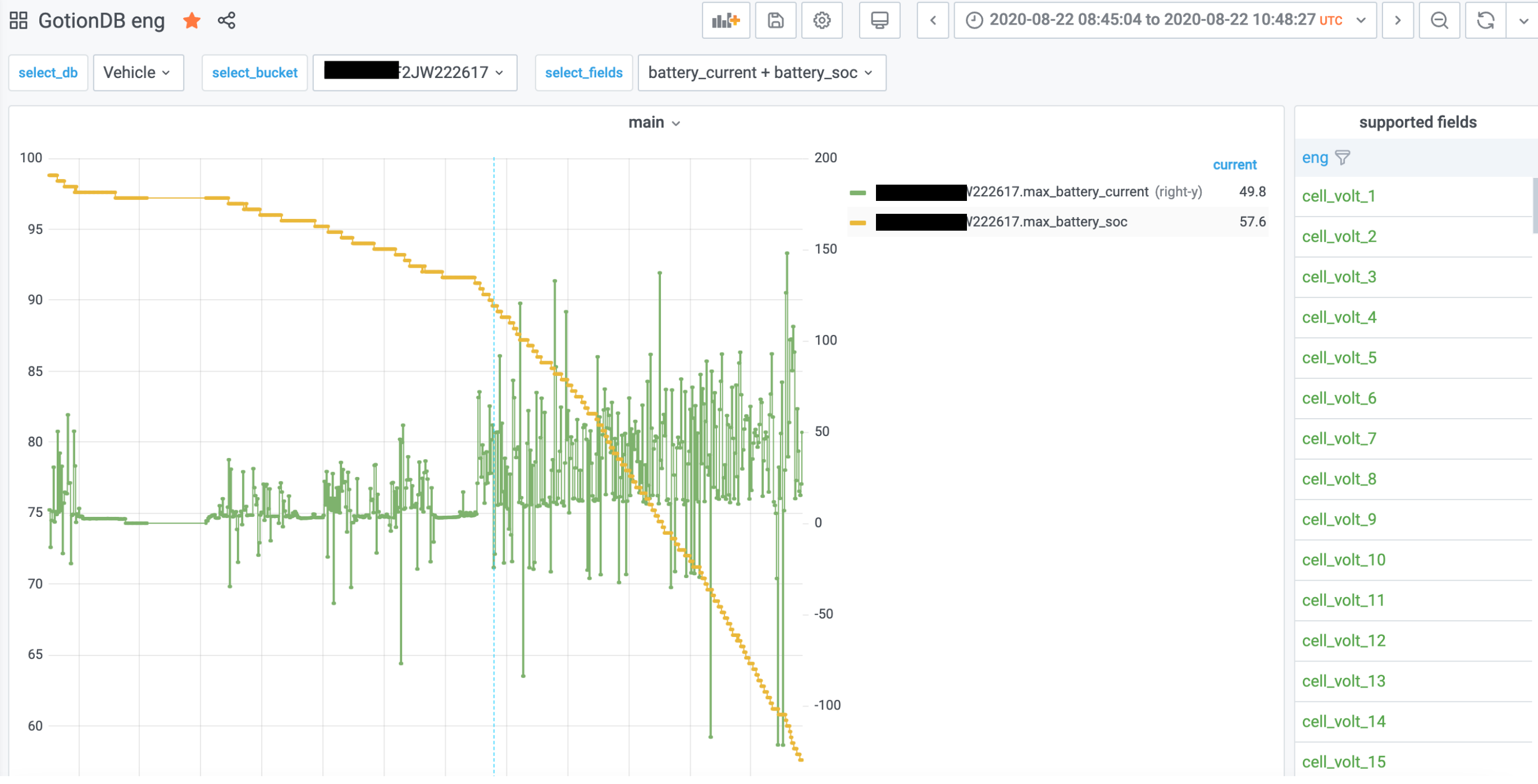}
\label{dashboa}
\caption{A typical dashboard for displaying battery data, developed by Gotion\cite{HowAWS}. }
\end{figure}

Most end-users are data analysts or operators who monitor EV/ESS in real-time. It is vital to have a responsive and interactive data visualization tool where essential data are displayed in real-time. Users can create a dashboard and add custom processing/query to explore statistical insights. Other features include: 1) adding a signal threshold, which can trigger quick alarms to the ESS site operator. 2) Options to trigger an ML pipeline from the frontend. Widespread data visualization tools are web-based, such as Grafana, Datadog, and Kibana. Figure~\ref{dashboa} depicts an example battery data display dashboard.

\subsection{Algorithms and Analytics}

With the data platform built, advanced algorithms that leverage big data and machine learning\cite{Lombardo2021ArtificialReality} can be applied to increase battery performance, safety and economy. Another interesting topic is the digital twin \cite{Li2020DigitalEstimation}. Based on sophisticated electrochemical modeling, the digital twin can give insight into the internal states of its physical twin. The battery cloud platform will need API (application programming interface) for popular programming languages, such as Python and Matlab\textregistered, based on the developers' preferences. It may also provide a more interactive computing platform like the Jupyter notebook/Lab, commonly used for data analytics. After the algorithms/analytics are developed, they should be optimized and incorporated into a data processing engine, such as Spark, Kafka, and Airflow. Life data of all the battery cells  are used to analyze the manufacture, assembly process, and facility to improve quality management. Similarly, these data can be used as references during battery second life application, recycling, and refurbishing, eliminating the need for extra testing/calibration.

\section{Onboard SOC Estimation with Cloud-trained ANN}

State of Charge (SOC) estimation is one of the essential functions of battery software. It has been researched extensively. There are mainly three different methods for SOC estimation. The commonly used, basic method is coulomb counting, which calculates the accumulated charge by current integral, given as

\begin{equation}
    z(t)=\frac{1}{C}\int_0^t{i(\tau)}d\tau + z(0),
\end{equation}

where $C$ is the battery capacity. This method is susceptible to accumulated error generated from $i(t)$ or data loss from $z(0)$. As such, estimation accuracy degrades if coulomb counting is used without correction over a prolonged period of time. The other two methods are model-based and data-driven. Both have self-correction features to correct SOC. The model-based approach utilizes a battery model, either ECM (equivalent circuit model) or electro-chemical model, to establish the connection between battery measurements, such as voltage, temperature, and current, and immeasurable internal states. Then an estimator, such as Kalman Filter, is applied to estimate the SOC. Because those batteries are highly nonlinear systems, modified Kalman filters such as extended Kalman filters and unscented Kalman filters are often used in practice. Model-based approaches require an accurate model. The model can be calibrated accurately by long-term cell and pack testing. But it is also prone to over-fitting, meaning it can not tolerate individual cell/pack variations. Making an accurate and well-generalized model is very challenging and time-consuming. Data-driven approaches range from simple voltage-based correction\cite{9699170} to deep neural networks\cite{Chemali2018State-of-chargeApproach}. More detailed reviews of data-driven SOC estimation methods are covered by \cite{Lombardo2021ArtificialReality,Hannan2017}. Most of these methods are resource-consuming and can not be easily applied to an onboard BMS. 

This section will present a data-driven SOC estimation method that fuses the onboard BMS with the battery cloud. A neural network is firstly trained with cloud battery data. The neural network is designed to reduce its computational and memory footprint to be fit into a micro-controller.

\subsection{Requirements Definition and Design}

\begin{figure}[t]
\centering
\includegraphics[width=\columnwidth]{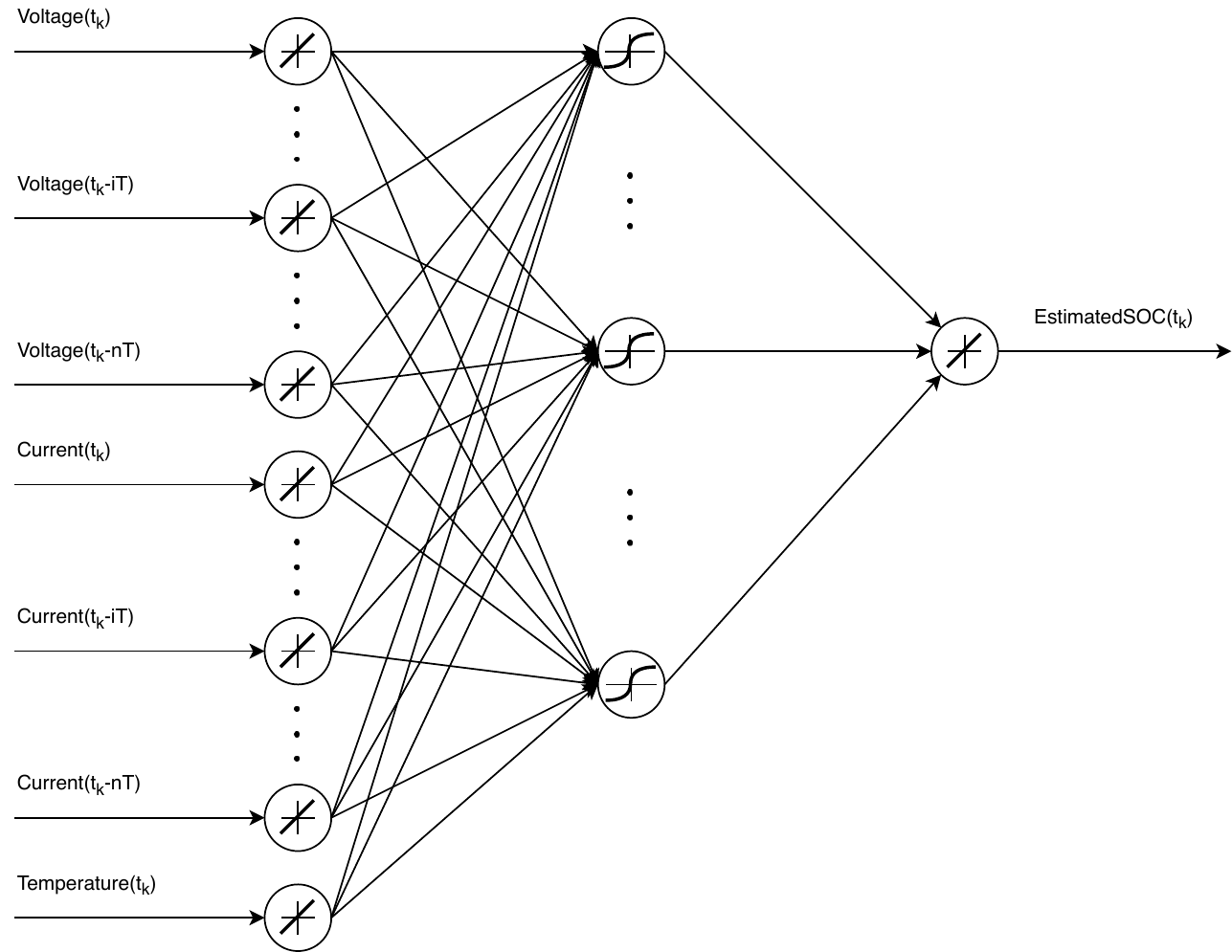}
\caption{The neural network model used for SOC estimation}
\label{ann_de}
\end{figure}

When designing SOC algorithms, there are several typical requirements to be considered. For example:

\begin{enumerate}
    \item SOC estimation should be 100\% when the battery is fully charged.
    \item SOC estimation should not change suddenly, including power cycles.
    \item SOC estimation should have a maximum error of less than 5\%.
    \item SOC estimation should have an average error of less than 3\%.
\end{enumerate}

The neural network presented is designed to meet requirements \#1,\#3, and \#4. Other requirements, like \#2 are implemented by a different software component, such as the SOC initialization function.

As depicted in Figure~\ref{ann_de}, the neural network includes an input layer, hidden layers, and an output layer. The inputs are measurable battery signals, including voltage, current, and temperature. Both present and historical measurements are used. Historical measurements are critical for the feed-forward ANN to infer the internal SOC of the battery, which is a dynamical system. 

\subsection{ANN Training with Cloud Data}

\begin{figure}[t]
\centering
\includegraphics[width=\columnwidth]{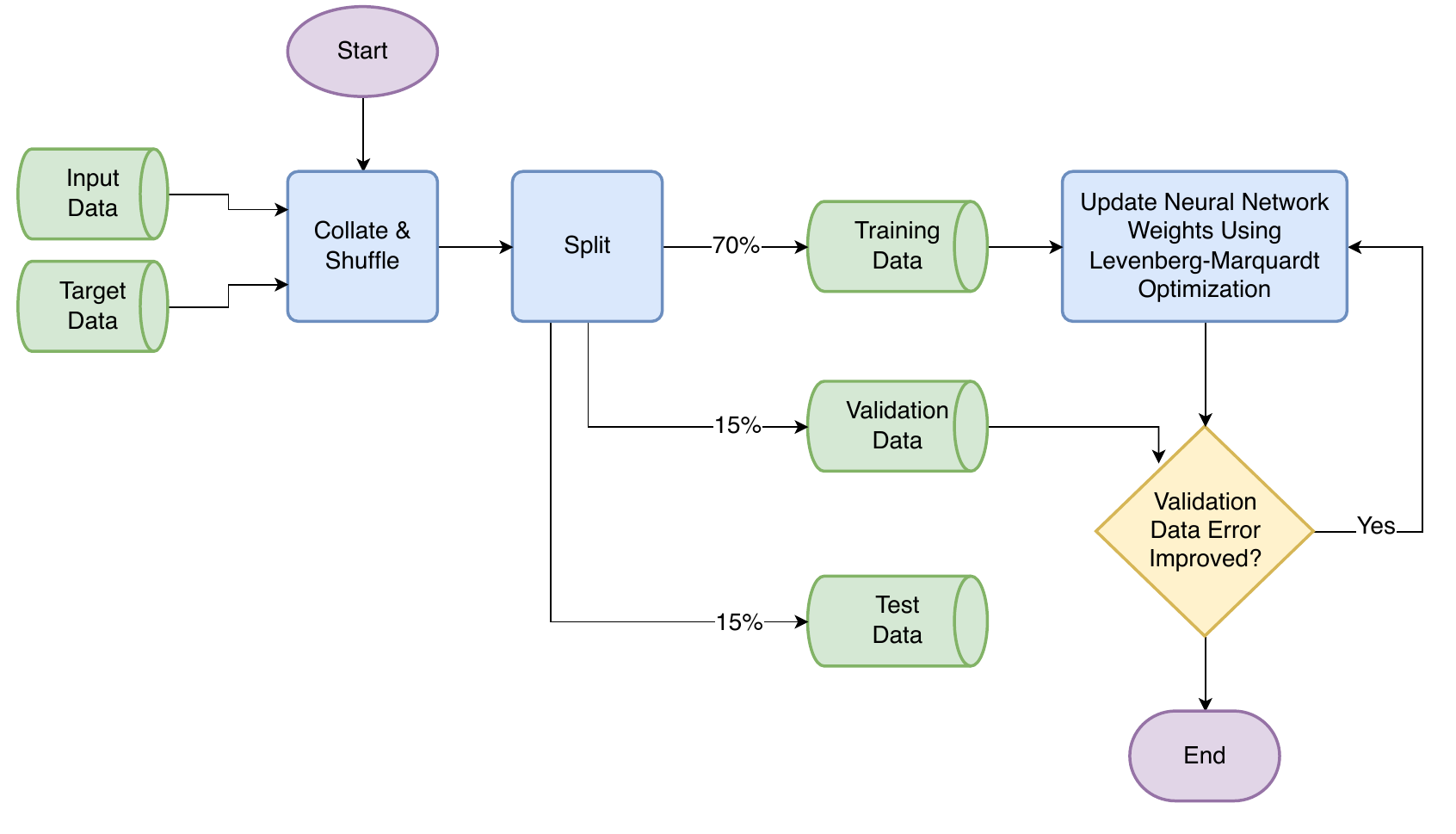}
\caption{The ANN training flowchart for SOC estimation}
\label{ann_tr}
\end{figure}

\begin{figure}[t]
\centering
\includegraphics[width=0.75\columnwidth]{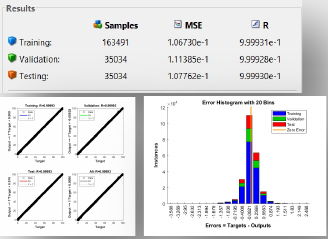}
\caption{ANN training results}
\label{ann_tr_rt}
\end{figure}

The ANN is developed using Matlab\textregistered~and Simulink\textregistered. Cloud battery data are fetched through Matlab\textregistered~API and used to train the neural network. DC charging data of NMC cells are used for training. The training data includes the cells being charged at a range of temperatures, including -10\textdegree, 0\textdegree, 25\textdegree, 40\textdegree, and 50\textdegree, during which the voltage and current signals are recorded. As depicted in Figure~\ref{ann_tr}, these data are split into training data (75\%), validation data (15\%), and test data (15\%). During the training, Levenberg-Marquardt is configured as the optimization algorithm. The training results are depicted in Figure~\ref{ann_tr_rt}. 

After acquiring the parameters, the ANN is implemented as Matlab\textregistered~code and integrated into the SOC software component, a Simulink\textregistered~model. Using the embedded coder, the Simulink\textregistered~model is converted to C code, integrated with other software components, and eventually become executable binaries.

\subsection{HIL and Vehicle Testing Results}

The ANN is first tested using the hardware-in-the-loop (HIL) system, during which basic functions of the software component and SOC accuracy are evaluated using cloud data. More importantly, as the ANN executes in the BMS real-time operating system (RTOS), the impact on CPU and RAM usage is evaluated. It is found that the ANN takes approximately 50 \textmu{s} of execution time. Its RAM usage is also small.

\begin{figure}[t]
\centering
\includegraphics[width=\columnwidth]{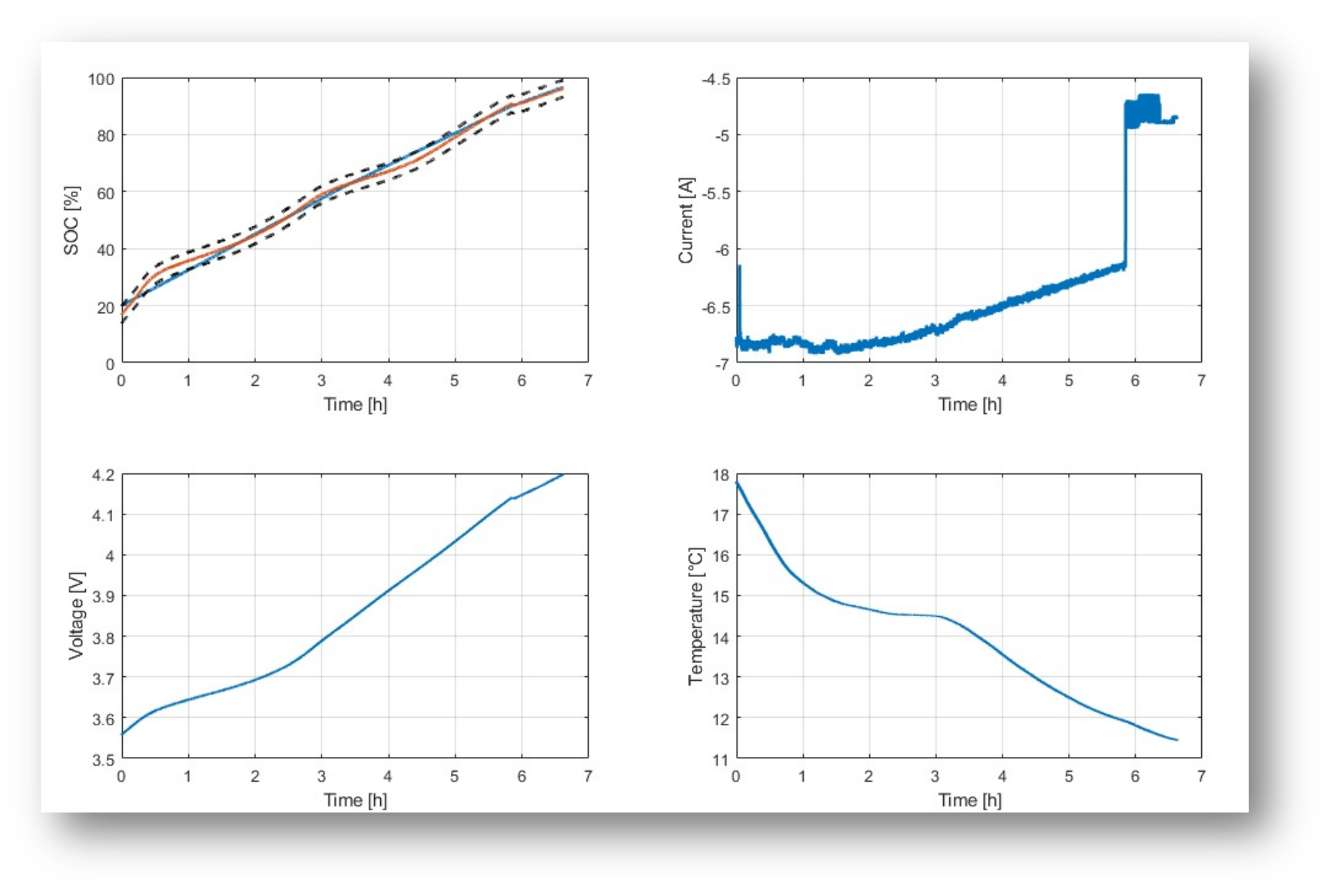}
\caption{Onboard Vehicle Test: AC charging}
\label{ann_ac}
\end{figure}

Finally, the algorithm is tested on a vehicle BMS. The ANN is deployed as a shadowing strategy in addition to existing software for several passenger EVs. To verify its robustness, the ANN is tested under the AC charge scenario to verify its robustness. Even though it is only trained with DC charge data, the ANN performed satisfactorily during AC. For example, Figure~\ref{ann_ac} depicts the SOC comparison, current, voltage, and temperature plots of one test. As shown in the SOC comparison plot, for most of the time, the true SOC (solid blue) falls into the +/-5\% bracket of the SOC estimation (solid red). The RMSE of all testing results is 1.9\%, well below the 3\% target.

\section{Online State-of-Health Estimation}
Li-ion batteries and many other secondary cells are subject to different degradation mechanisms that lead to a loss of removable energy or power, which lead to a decrease in range and acceleration for battery electric vehicle \cite{Birkl.2017}. For that reason, it is essential to monitor the state of health of Li-ion batteries. The degradation mechanisms are briefly discussed and organized into three categories in the following section. Furthermore, the concept of the end-of-life (EOL) and the SOH\textsubscript{C} will be presented. Conclusively, an overview of SOH\textsubscript{C} estimation methods will be given, with a closer look at the  voltage analysis (DVA) and the incremental capacitance analysis (ICA).

\subsection{Degradation Mechanisms and Modes of Li-Ion Batteries}
\begin{figure}[htbp]
    \centering
    \includegraphics[width=\textwidth]{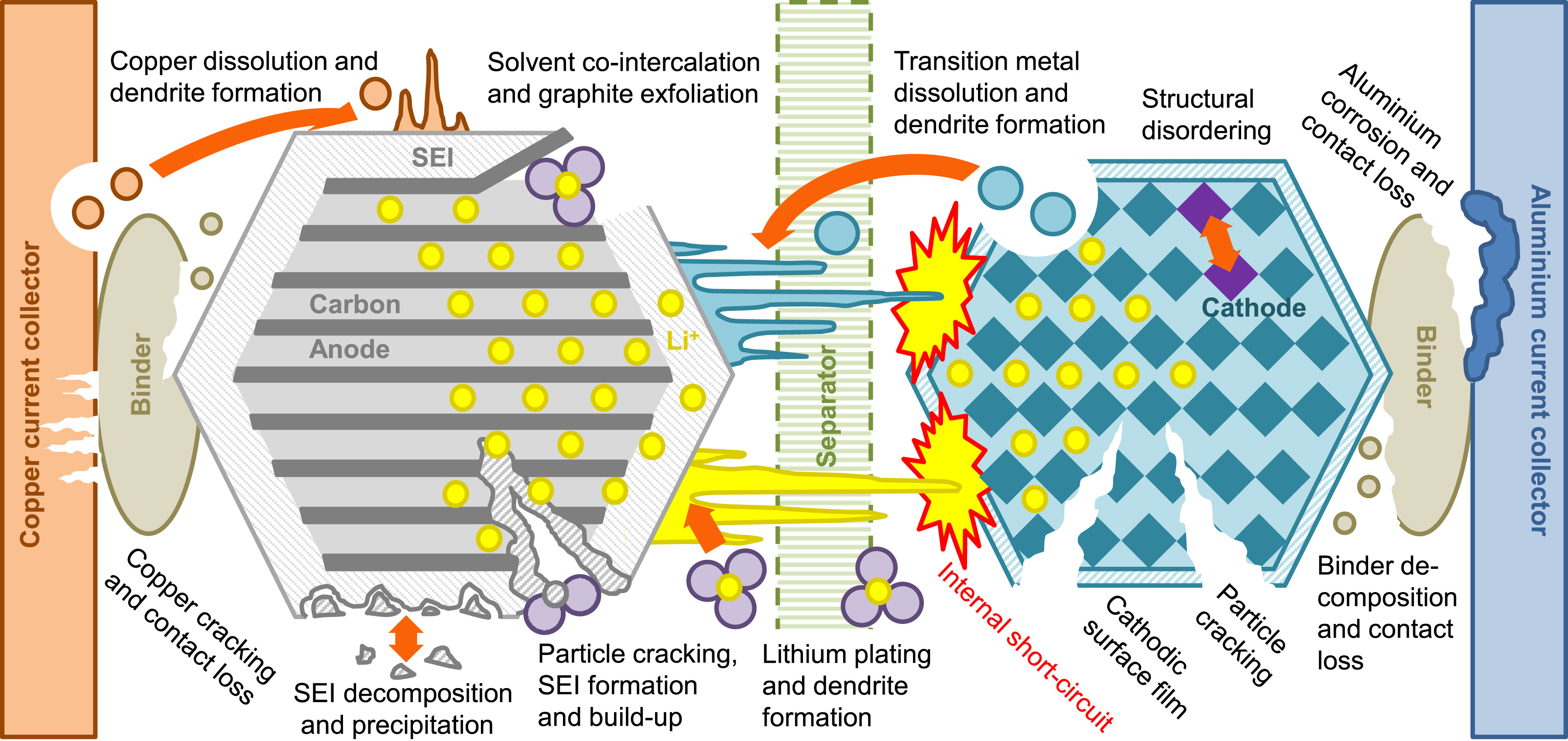}
    \caption{Degradation mechanisms in Li-ion cells adopted from Birkl et al. \cite{Birkl.2017}}
    \label{fig:agingmechanisms}
\end{figure}

The components of a Li-ion battery are subject to different degradation mechanisms. In general, the degradation mechanisms can be classified into three modes \cite{Birkl.2017, Vetter.2005, Dubarry.2012}:
\begin{itemize}
    \item \textbf{Loss of Lithium Inventory} (LLI): The Li-ions are consumed by side reactions and are no longer available for intercalation/deintercalation in the anode and cathode.
    \item \textbf{Loss of active material of the anode} (LAM\textsubscript{A}): The anode active material is no longer available for lithium intercalation due to particle cracking and loss of electrical contact or active areas being blocked by solid surface layers.
    \item \textbf{Loss of active material of the cathode} (LAM\textsubscript{C}): The cathode active material is no longer available for lithium intercalation due to structural failure, particle cracking, or loss of electrical contact.
\end{itemize}
These modes can be attributed to different degradation mechanisms in the components of a Li-ion cell. In the following, the primary degradation mechanisms on the components of a Li-ion cell are named and briefly explained. An overview of these degradation mechanisms is presented in the Figure~\ref{fig:agingmechanisms}.

\paragraph{Anode}
The main degradation modes at the negative electrode are Lithium Plating and the formation of solid electrolyte interphase (SEI).
Lithium plating is a commonly recognized and inherently damaging degradation mechanism in Li-ion batteries, which describes the deposition of lithium metal on the surface of the anode as soon as the anode potential exceeds the threshold of 0V (vs. Li/Li+) \cite{Kabir2017DegradationReview}. On the other hand, the solid electrolyte interphase (SEI) is a protective layer on the surface of the anode particles due to the decomposition of the electrolyte, which is formed mainly during the first cycles \cite{Verma.2010}. Both degradation modes are the main contributor for LLI and LAM\textsubscript{A} \cite{Birkl.2017}.

\paragraph{Cathode}
On the other side of the battery, the degradation modes of the cathode are still growing in interest and therefore not thoroughly documented yet. It is considered that structural changes and mechanical stress are the main contributors to LLI and LAM\textsubscript{C}. Due to various cathode materials, the Li-ion battery suffers from different side reactions based on the cathode material composition. For example, an Mn-based cathode is more prone to the dissolution of the active material due to Mn dissolution. In contrast, the degradation of the LFP cathode is more likely to be defined by Fe dissolution, which generates HF as a by-product and attacks the surface of the cathode particles \cite{Kabir2017DegradationReview}.

\paragraph{Separator, Electrolyte and Current Collectors}
The separator, electrolyte and current collectors also suffer from various degradation mechanisms. Although the porous separator of a Li-ion cell is electrochemically inactive, the separator can negatively affect the performance of the cell. According to various studies, deposits from the electrolyte decomposition can clog the pores of the separator. This can increase the impedance and also can reduce the accessible active surface of the electrodes (LAM\textsubscript{A} and LAM\textsubscript{C}, respectively) \cite{Vetter.2005, Aurbach.1999}.
On the other hand, the electrolyte is involved in side reactions that lead mainly to the formation of a surface film on the negative electrode, but also partly on the positive electrode. Electrolyte oxidation at the cathode does not directly affect any of the three modes of degradation, but it does cause re-intercalation into the active material, also known as self-discharge of the cell. In contrast, electrolyte reduction at the anode results in a loss of cyclable lithium, which leads to a capacity loss. \cite{Keil.2017, Broussely.2005}.
The current collectors of a Li-ion cell suffer mainly from two degradation mechanism. First, the current collectors can corrode electrochemically. This occurs particularly at the aluminum collector of the positive electrode when acidic species, such as HF, are present \cite{PankajArora.1998}. The copper current collector of the negative electrode may dissolve during deep discharge when the anode potential increases to $\SI{1.5}{V}$ in reference to \ce{Li/Li^{+}} \cite{Han.2019}. Second, the current collector foils may deform due to mechanical stress. This can disrupt the contact between the electrodes and the separator so that specific active areas can no longer participate in the intercalation process of Li-ions in to the electrodes, which causes a loss in capacity \cite{T.Waldmann.2014}.

\begin{figure}[htbp]
    \centering
    \includegraphics[width=\textwidth]{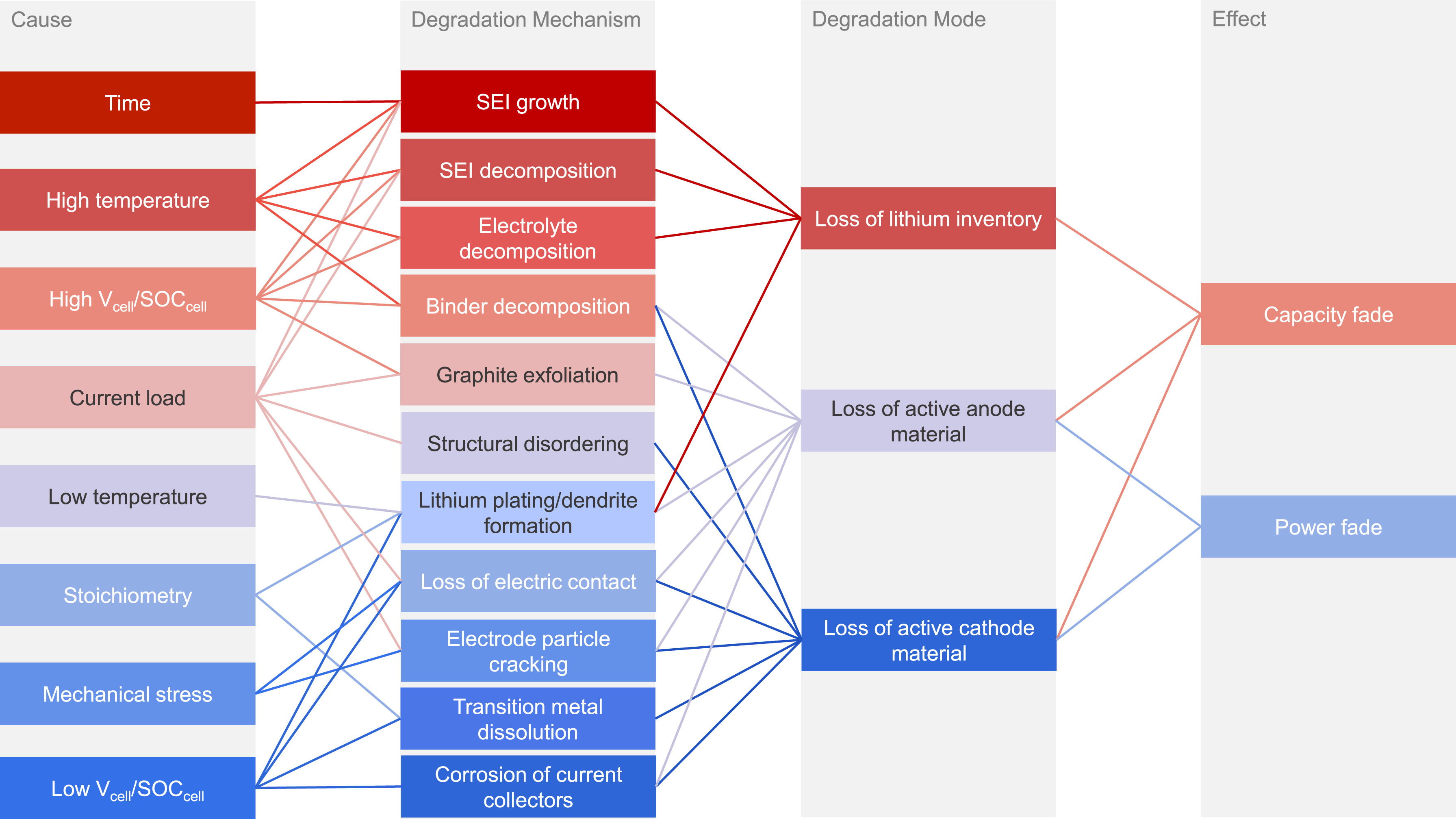}
    \caption{Overview of different degradation mechanisms, their cause and their effect on the performance of the Li-ion cell adopted from Birkl et al. \cite{Birkl.2017}}
    \label{fig:overviewAM}
\end{figure}

Based on the Figure~\ref{fig:overviewAM}, it can be seen that all degradation modes can be organized by their effect on the electric characteristics of a Li-ion battery, the capacity and power fade. For further elaborations, we will focus primarily on the capacity fade.

\subsection{State-of-Health \& End-of-Life}
Due to the degradation mechanisms, Li-ion batteries have a limited lifetime. The end of life (EOL) of a Li-ion battery is reached when the battery can no longer provide the power or energy intended for its application \cite{Korthauer.2013}. However, as of today, there is no uniform standard that defines a clear EOL criterion for Li-ion batteries in the new energy industry \cite{Keil.2017}. The USABC consortium is the only one to define two EOL criteria in its manual of test procedures for electric vehicle batteries. According to this manual, the EOL of a Li-ion battery is reached when:
\begin{itemize}
    \item the net capacity delivered is less than $80\%$ of the rated capacity C\textsubscript{N} or,
    \item the peak capacity is less than $80\%$ of the rated capacity at a DOD of $80\%$ \cite{Hunt.1996}.
\end{itemize}
Also, in many publications, the EOL for the Li-ion-based traction battery of a BEV at a capacitive aging condition of $SOH\textsubscript{C} \leq 80\%$ assumed \cite{Wood.2011, V.Marano.2009, YanchengZhang.2011}. The capacitive aging state SOH\textsubscript{C} can be calculated using equation \ref{equ:SOH_C}. Here, C\textsubscript{N} corresponds to the nominal capacity of the Li-ion battery, and Q\textsubscript{dis,max} corresponds to the maximum charge quantity that can be removed from a Li-ion battery, which is also known as the net capacity. 
\begin{equation}
    \label{equ:SOH_C}
    SOH_C = \frac{Q_{dis,max}}{C_N}\cdot 100\%
\end{equation}

\subsection{Advanced Online SOH-Estimation Methods}
The capacitive aging state SOH\textsubscript{C} has an impact on two critical factors of a BEV, the maximum range and the charging time during fast charging. Based on the capacitive aging state SOH\textsubscript{C}, the maximum range can be predicted to the driver, and the fast charging function can be adjusted to find the optimal compromise between minimum charging time and damage of the anode by lithium plating \cite{Tomaszewska.2019}. Therefore, an online SOH\textsubscript{C} estimation is essential for automotive applications.

\begin{figure}[htbp]
    \centering
    \includegraphics[width = 0.75\textwidth]{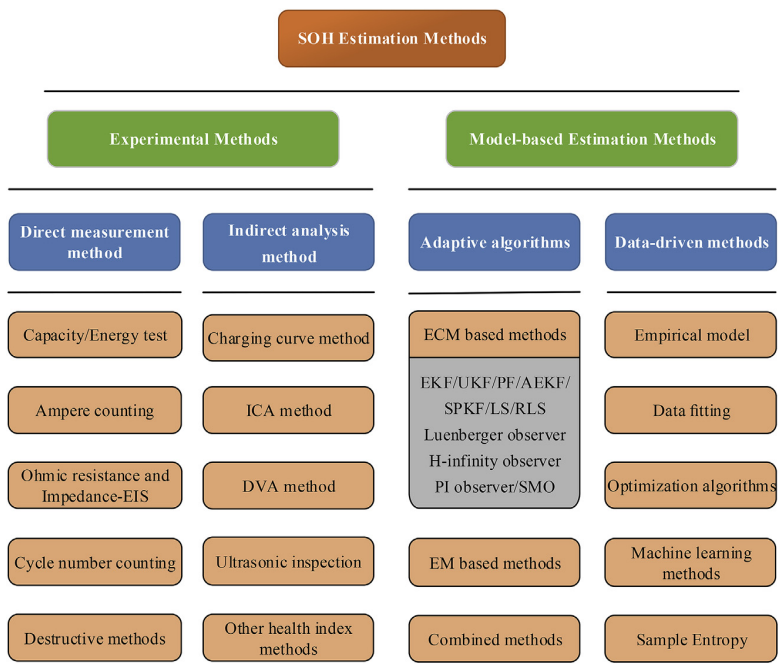}
    \caption{Methods for SOH\textsubscript{C} estimation adopted from Xiong et al. \cite{Xiong.2018}}
    \label{fig:soh_overview}
\end{figure}

\subsubsection{Methods}
There are several options for determining the capacitive aging state, which can first be divided into experimental and model-based methods, respectively. As shown in Figure~\ref{fig:soh_overview} \footnote{This figure was published in Journal of Power Sources, 405, Rui Xiong, Linlin Li, Jinpeng Tian, Towards a smarter battery management system: A critical review on battery state of health monitoring methods, 18-29, Copyright Elsevier (2018).}, these categories can be subdivided into further subcategories. For example, experimental methods can be divided into direct methods, such as capacity measurement (coulomb counting), and indirect methods, such as differential voltage analysis. In contrast, model-based methods can be divided into SOH\textsubscript{C} determination based on adaptive battery models or data-driven methods.

At this point, the most popular methods regarding the capacitive aging condition SOH\textsubscript{C} are mentioned and briefly discussed. More detailed summaries about methods for SOH\textsubscript{C} estimation were given by Berecibar et al. \cite{Berecibar.2016b} and Xiong et al. \cite{Xiong.2018}.

\paragraph{Direct measurements} are the most straightforward method to determine the SOH\textsubscript{C}. One prevalent method is the direct measurement of the current battery capacity. However, this method requires an enormous expenditure due to the low charging current during the capacity measurement, which is why this method is only used for R\&D purposes.

\paragraph{Model-based estimation methods} utilize algorithms like Kalman Filter or Neural Networks to model the battery cell parameters. These methods achieve relatively high accuracy and can be implemented in a Cloud-BMS. Yet, these algorithms require a high development effort. For example, the accuracy of the Kalman filter is highly dependent on the accuracy of the applied battery model, whereby high accuracy is only achieved with complex battery models. Also, training a neural network requires a large amount of data, which can only be generated by cost-intensive testing of battery cells. In addition, these algorithms need extensive validation. For example, the neural network is considered a black box, and the output can not be generally predicted based on unexpected input data.

\paragraph{Indirect analysis methods} utilize various battery parameters to correlate the capacity fade with various features of the Li-ion battery. For example, the charge curve can characterize the SOH\textsubscript{C} of the battery as it changes throughout the battery degradation. Constant current followed by constant voltage with current limiting (CCCV) charging mode is commonly used for batteries. Eddahech et al.\cite{Eddahech2014DeterminationPhase} developed a method for SOH\textsubscript{C} estimation using the CV stage as a health indicator. Since minimal intrinsic information about the battery can be obtained directly from the voltage curves, Dubarry et al. \cite{Dubarry2009IdentifyCell, Dubarry2011IdentifyingCells}, for example, used electrochemical characterization and analysis techniques, incremental capacitance analysis (ICA), and differential voltage analysis (dV/dQ) (DVA). These methods are often applied in laboratories since a low current rate is required to record these differential curves. However, due to the increasing energy of the battery packs and the lower power of AC charging, it is also possible to record the differential curves during an AC charging process in a BEV. For this reason, the basics of the DVA and ICA and their correlation with capacity fade will be discussed in more detail in the following section.

\begin{figure}[htbp]
    \centering
    \includegraphics[width = 0.75\textwidth]{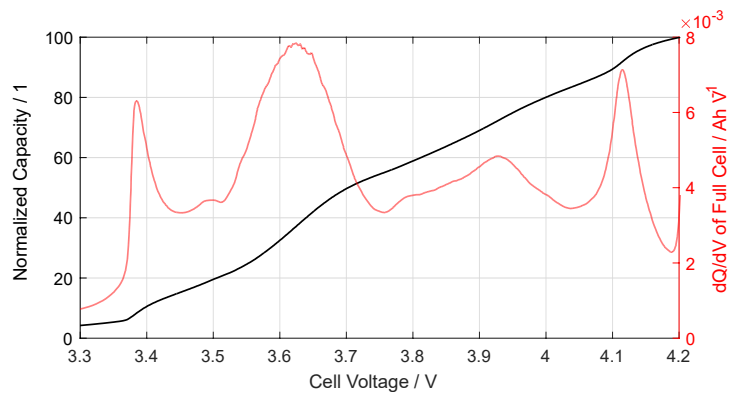}
    \caption{The differential curve of ICA during a CC charge with C/10 of a Li-ion battery consisting of a graphite anode and NMC cathode.}
    \label{fig:dqdv}
    \centering
\end{figure}

\subsubsection{DVA/ICA-based SOH-Estimation Method}
As mentioned before, it is essential to estimate the SOH\textsubscript{C} of battery packs in BEV. In the following section, a DVA and ICA-based estimation method is introduced. Therefore, the DVA and ICA will be presented and discussed in detail to present a simple SOH-estimation implementation, which could be realized on a cloud platform.

The DVA and ICA are commonly known analysis methods for Li-ion batteries in laboratories. The IC curves can be calculated from equation~\ref{equ:dqdv} during a low-current charging or discharging process. The division of an infinitesimal charge change due to the charge/discharge current by the resulting voltage change is calculated during the charge/discharge process. This process converts the low-slope regions of the OCV curve, also known as voltage plateaus of the two-phase transition, into detectable IC peaks. Another method, is the differential voltage analysis (dV/dQ) (DVA). The DV curves can be calculated by the reciprocal of the IC curve as shown in equation~\ref{equ:dvdq}. The distance between two peaks of the DV curve represents the amount of charge involved in the two-phase transition, so it is easier to analyze the capacity degradation quantitatively using the DV curves \cite{Han.2014}.

\begin{equation}
\frac{dQ}{dV} \approx \frac{Q(t) - Q(t+\Delta t)}{V(t) - V(t+\Delta t)}
\label{equ:dqdv}
\end{equation}
\begin{equation}
\frac{dV}{dQ} = (\frac{dQ}{dV})^{-1} \approx \frac{V(t) - V(t+\Delta t)}{Q(t) - Q(t+\Delta t)}
\label{equ:dvdq}
\end{equation}

The result of calculating the IC curve during a low current rate charging process is shown in Figure~\ref{fig:dqdv}. It should be noted that the DV curves can be represented using the half-cell potentials due to the superposition behavior of the anode and cathode (see Equation \ref{equ:dvdq}). Thus, the peaks and valleys of the DV curve can be assigned to the anode and cathode, respectively.

% \begin{figure}[t]
%     \centering
%     \includegraphics[width = 0.75\textwidth]{pics/DVA_VZ_HZ_SoC.eps}
%     \caption{Full CC charge of a Li-ion battery with a graphite anode and NMC cathode with C/10 (a) Voltage curve of the full cell calculated from the electrode potentials \& electrode potentials of the half cells vs Li/Li+ over the entire SoC range (b) Difference curves of the DVA of the half cells and the full cell calculated from them.}
%     \label{fig:dvdq}
%     \centering
% \end{figure}

\begin{figure}[ht]
    \centering
    \includegraphics[width = 0.85\textwidth]{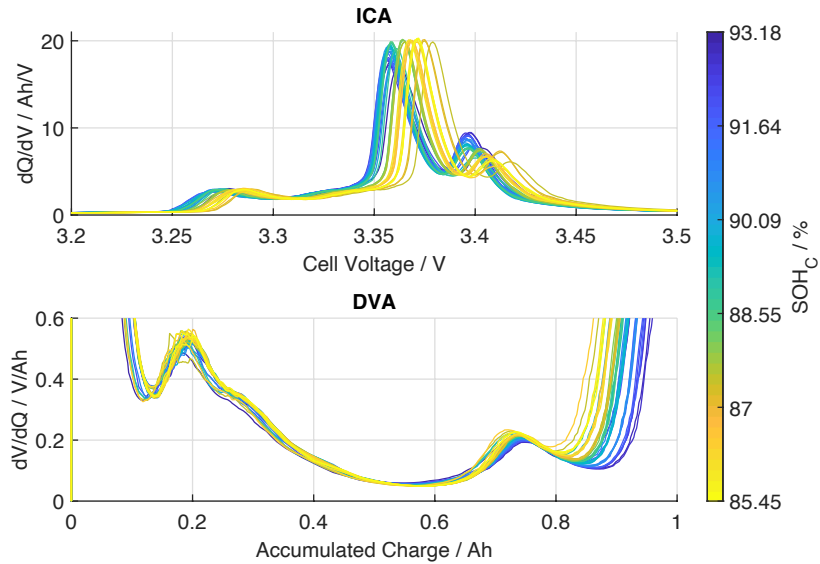}
    \caption{The course of the differential curves of the (1) ICA and (2) DVA during a C/2 of a cyclically aged Li-ion battery (graphite anode/LFP cathode).}
    \label{fig:CyclicAgingDVAICA}
    \centering
\end{figure}

As mentioned before, Li-ion batteries suffer from various degradation mechanisms, which lead to LLI, LAM\textsubscript{C} and LAM\textsubscript{A}. Due to these degradation modes, a change in the DV and IC curves can be observed. Figure~\ref{fig:CyclicAgingDVAICA} shows the shift of the DV and IC curve due to cyclic aging. Based on the change of these features, the SOH\textsubscript{C} can be estimated by correlation, for example, the distance between two peaks in the DV curve with the capacity fade of the Li-ion battery. Another possible feature is the height or depth of the peaks or valleys of the IC curve, which also shift throughout the ongoing degradation of battery materials.

\begin{figure}[ht]
    \centering
    \includegraphics[width = \textwidth]{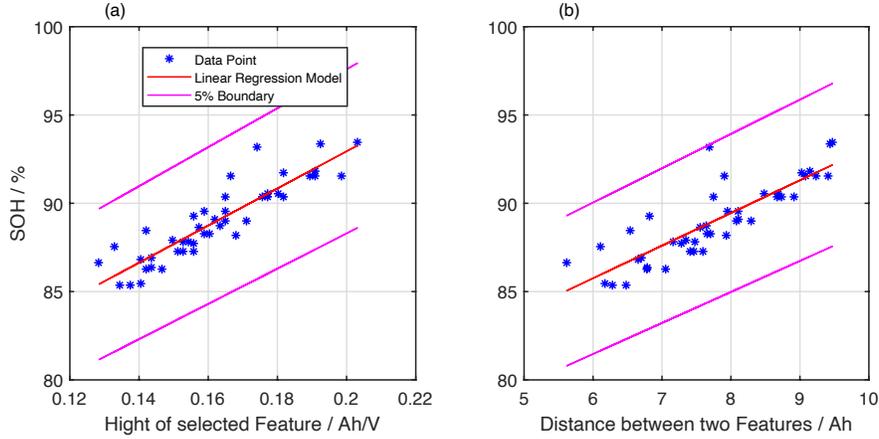}
    \caption{Correlation of selected features with Capacity Fade: (a) Height of selected Feature (b) Distance between two features}
    \label{fig:LinRegressionSelectedFeatures}
    \centering
\end{figure}
In order to implement a  DVA/ICA-based SOH estimation on a Battery Cloud the following workflow should be included:
\begin{enumerate}
    \item The platform monitors the typical battery cell parameters, voltage, current, and temperature.
    \item Whenever the battery is charging, it determines if the charging data has satisfied feature detection based on several conditions. The conditions include the C rates, amount of charge, and so on.
    \item If the conditions are met, proceed with the following steps. Otherwise, abort and watch for the next window.
    \item Calculate and filter the differential curves (dV/dQ) based on the measurements.
    \item Apply feature detection algorithm, i.e., a peak detection algorithm, to extract the features. Based on the scenarios, different features may be extracted and used. Since the features themselves do not indicate the SOH\textsubscript{C}, they will be further processed.
    \item Apply a mapping function that relates the features with the SOH\textsubscript{C}. Typically, the reference is represented by a Look-Up-Table (LUT) that is based on the correlation between features and SOH\textsubscript{C}, extracted from existing cyclic aged battery data.
\end{enumerate}

% Observing Figure~\ref{fig:CyclicAgingDVAICA} it can be seen that the features can only be detected during specific SOC windows. Therefore, the estimation rate with this described method is limited. To overcome this issue, Fend et al. \cite{Feng.2019} integrated an SVM to predict the curve in case the charging process was terminated before the feature was detected. This SVM could be easily realized on the Cloud-BMS, whereas this algorithm would possibly reach its limit on a conventional controller of a standard BMS.

As depicted in Figure~\ref{fig:LinRegressionSelectedFeatures}, the real SOH\textsubscript{C} has strong correlations with DVA and ICA features. For example, the distance of two features or the hight of a feature. The correlations also depend on the temperature. Higher charging currents will affect the estimation accuracy. However, this method can generally achieve 5\% SOH accuracy when charging at C/2 or less.

% \begin{itemize}
%     \item introduce DVA/ICA
%     \item Explain in block diagram roughly the principle
%     \item ICA-Calc \textrightarrow Filter \textrightarrow DVA-Calc \textrightarrow Feature Detection \textrightarrow Feature Calculation \& Plausibility        Check \textrightarrow 2D-LUT
%     \item 2D-LUT generated from cells which have been cycled at various C-rate and temperatures.
%     \item MBD implementation in Simulink \textrightarrow C-code generation
% \end{itemize}

% \subsection{Conclusion and Outlook}

\section{Cloud-based Thermal Runaway Prediction }
\subsection{Cause and Effects of Thermal Runaway}
One significant disadvantage of batteries is the narrow operating temperature range. The safety and stability of the battery cells are dependent on keeping interior temperatures under certain limits. A thermal runaway can occur if the temperature surpasses the critical level, killing the battery or, even worse, causing a fire. Thermal runaway is a chain reaction that can be very difficult to stop once it has begun within a battery cell. During a thermal runaway, the temperature rises incredibly fast (milliseconds), and temperature can be higher than 752\textdegree{F}/400\textcelsius. At such elevated temperatures, electrolytes in the battery cell can be vaporized and combustive when exposed to oxygen. Such battery fires are hard to extinguish with conventional ways. 

\begin{figure}[t]
\centering
\includegraphics[width=0.6\columnwidth]{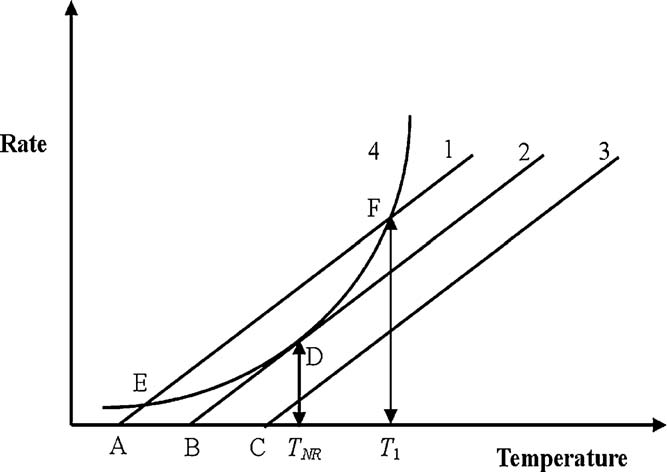}
\caption{Thermal Runaway explanation based on heat generation and dissipation models\cite{Wang.2012}}
\label{fig_tr}
\end{figure}

The heat generated by the electro-chemical reactions is critical as it can lead to thermal runaway. The heat generation is caused by chemical/electro-chemical reactions and joule heating inside the battery. Radiation and convection dissipate heat to the surroundings. The process of thermal runaway can be explained by the plot Figure~\ref{fig_tr}. The heat generation because of an exothermic reaction assuming Arrhenius law, an exponential function, is shown in curved line 4. In comparison, the heat dissipation is represented by straight lines, which follow Newton's cooling law at different coolant temperatures. For the lithium-ion battery, curve 4 is the combined results of reactions in the cell during the thermal runaway process and the energy balance between the heat generation. Heat dissipation is shown as the following equation Eqn. (\ref{eqn:tr})

\begin{equation}
\frac{\partial\left(\rho C_{p} T\right)}{\partial t}=-\nabla(k \nabla T)+Q_{\mathrm{ab}-\mathrm{chem}}+Q_{\mathrm{joul}}+Q_{\mathrm{S}}+Q_{\mathrm{P}}+Q_{\mathrm{ex}}+\cdots,
\label{eqn:tr}
\end{equation} where $\rho\left(\mathrm{g} \mathrm{cm}^{-3}\right)$ is the composite/average density of the battery, $C_{p}\left(\mathrm{Jg}^{-1} \mathrm{~K}^{-1}\right)$ the composite/average heat capacity per unit mass under constant pressure, $T(\mathrm{K})$ the temperature, $t(\mathrm{s})$ the time, $k\left(\mathrm{W} \mathrm{cm}^{-1} \mathrm{K}^{-1}\right)$ the thermal conductivity. $Q_{\mathrm{ab}-\mathrm{chem}}$ the abuse chemical reaction in the battery, $Q_{\text {joul}}$ Joule heat, $Q_{s}$ the entropy heat, $Q_{P}$ the overpotential heat, and $Q_{e x}$ the heat exchange between the system and the ambient.

Generally, thermal runaway can be triggered by various types of abuse in a battery shown in Figure~\ref{fig_ab} \cite{Feng2018review}, including:

\begin{figure}[t]
\centering
\includegraphics[width=0.75\columnwidth]{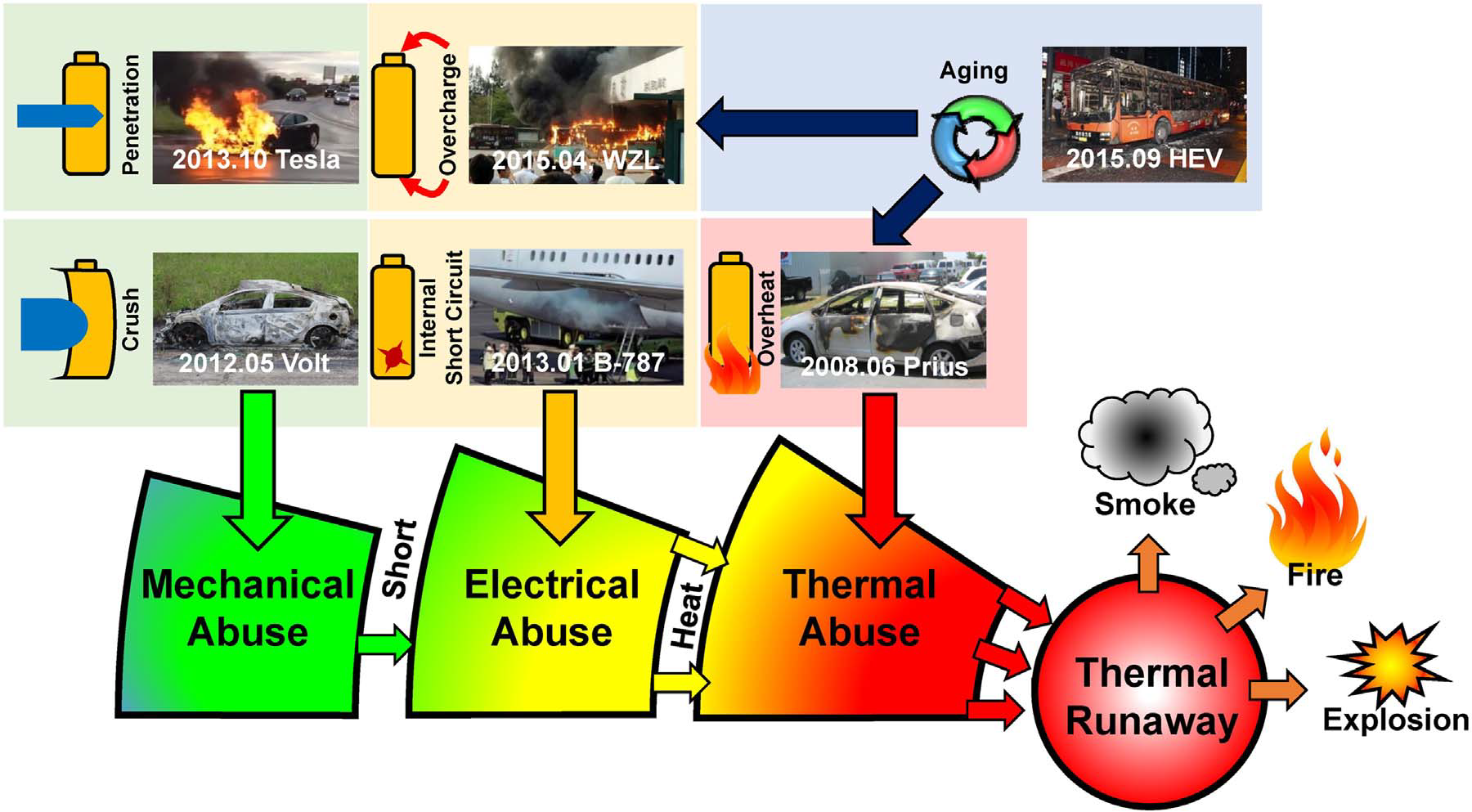}
\caption{Abuses that cause thermal runaway \cite{Feng2018review}}
\label{fig_ab}
\end{figure}

\paragraph{Internal short circuit} Internal short circuit caused by physical damage to the battery or poor battery maintenance. 

\paragraph{Mechanical abuse} Vehicle collision
and consequent crush or penetration of the battery pack are the typical conditions for mechanical abuse.

\paragraph{Electrical abuse} 
\begin{itemize}
    \item Overcharging: the voltage that exceeds the maximum safety operation voltage range will damage the battery and lead to thermal runaway. Because of the extra energy filled into the battery during overcharge, the overcharge-induced TR can be more severe than other abuse conditions.
    \item Rapid charging can lead to excessive currents, therefore, causing thermal runaway
    \item External short circuit: External short circuit happens when the electrodes with voltage difference are connected by conductors, which could also kick off the TR chain reaction. 
\end{itemize}

\paragraph{Thermal abuse}
\begin{itemize}
    \item Over/Under temperatures: either the low or high side of the safety ranges degrades battery health, leading to irreversible damage that may eventually trigger the TR reaction.
    \item  Contact loose of the cell connector can lead to overheating.
\end{itemize}

\subsection{Methods for Thermal Runaway detection}
 For the typical applications of batteries, including micro-grids and Electric Vehicles, they are connected and packed in modules and packs. Suppose one or a few batteries experience thermal runaway due to the limited space for heat exchange. In that case, the heat will rapidly go up, leading to thermal runaway propagation among all surrounding batteries. Therefore, it's essential to detect thermal runaways at an early stage to ensure operation safety. Lithium-ion batteries may experience a voltage and current anomaly, a temperature rise, or a gas venting during a thermal runaway process. Those are the indicators that can be detected at the early stage of thermal runaway to ensure the operation safety of batteries\cite{Liao.2019}. Methods of thermal runaway detection include:

\paragraph {Terminal voltage} The terminal voltage can be detected by using voltage sensors within the battery management system.

\paragraph {Mechanical deformation} Mechanical deformation can be detected by creep distance sensors.
\paragraph {Internal temperature} Since the core temperature directly represents the thermal condition within batteries, it can be either:
\begin{itemize}
    \item measured by temperature sensor inserted in the batteries
    \item estimated in terms of the measured surface temperature of batteries 
\end{itemize}
\paragraph {Gas component} Some gas components can be identified during the thermal runaway process, such as Carbon monoxide, hydrocarbons, and Hydrogen. Gas sensors like thermal conductivity detectors (TCD) can be used for this purpose.

\subsection{Data-driven Thermal Anomaly Detection}

Here we give a cloud-based and data-driven method for detecting battery thermal anomalies\cite{9576348}. Because that this method is based on the measurements' shape-similarities, which is less affected by cell deterioration or environmental variation, it is robust to battery aging or environment variations. As a result, this method can be applied to different battery configurations. The shape-based distance measurement handles the asynchronous data issue. It also needs very little reference data. This method is based on K-shape clustering\cite{Paparrizos2015} 

\begin{equation}
    SBD(\vec{x},\vec{y})\ =\ 1- \underset{\omega}{max}{\left(\frac{CC_\omega\left(\mathbf{x},\mathbf{y}\right)}{\sqrt{R_0\left(\mathbf{x},\mathbf{x}\right)R_0\left(\mathbf{y},\mathbf{y}\right)}}\right)},
    \label{k-shape}
\end{equation} where \(\vec{x},\vec{y}\) are two time-series measurements that used for comparing similarity, and  \(R_0\) the Rayleigh Quotient.

\begin{figure}[]
\centering
\includegraphics[width=0.6\columnwidth]{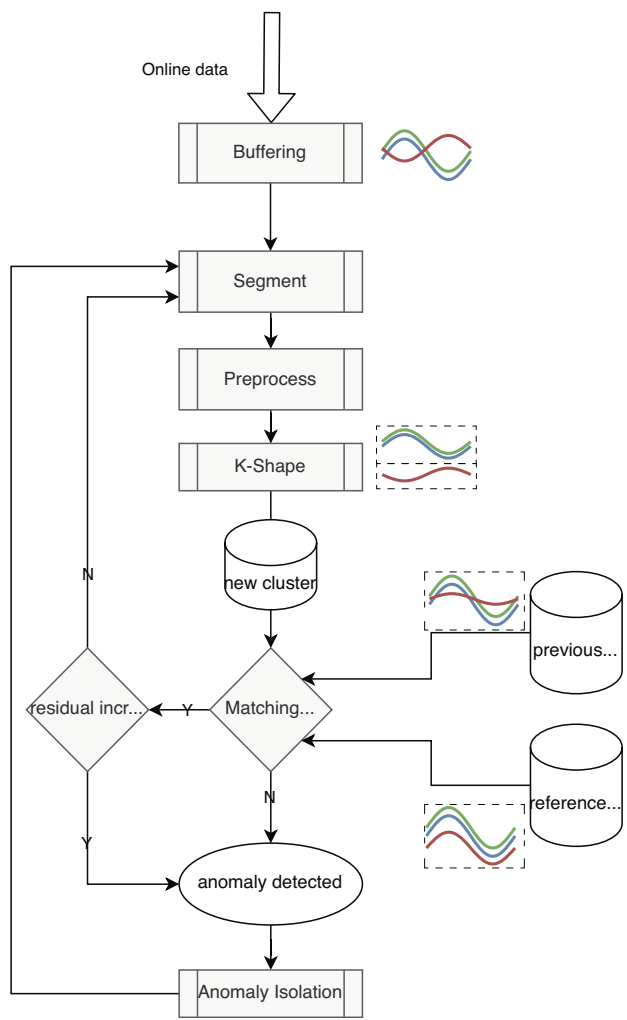}
\caption{Flowchart for the thermal anomaly detection algorithm}
\label{fig_sim}
\end{figure}

\subsubsection{Workflow} 

As depicted in Figure~\ref{fig_sim}, the proposed anomaly detection method contains the following steps. At first, data is continuously buffered and segmented. During the pre-processing stage, invalid/faulted data points are removed. Signals are normalized. Segments with static signals are filtered out.During the Anomaly Confirmation stage, the K-shape algorithm is applied to each segment for the distances ($SBD(x_i,c_j)$) for each cluster. Two criteria are used for determining anomaly. 1) When at least one of the measurements change the membership. 2) When no change was found in cluster membership, we check for any noticeable increase in the fitting errors. During each iteration, the \textit{i}th cluster is compared to the reference cluster, which captures the accumulated, long term changes that result from anomalies caused by gradual deterioration. Such as thermal anomalies caused by increased battery impedance. It's also compared to the predecessor for anomalies that developed abruptly, such as short-circuit. In the final stage of Anomaly Isolation, we use the change of membership or increase in fitting error to isolate the signals that have caused the anomaly.

\subsubsection{Case Study} 

\begin{figure*}[]
\centering
      \subfigure[two months of temperature signals, over-temperature highlighted]{\includegraphics[width=0.9\textwidth]{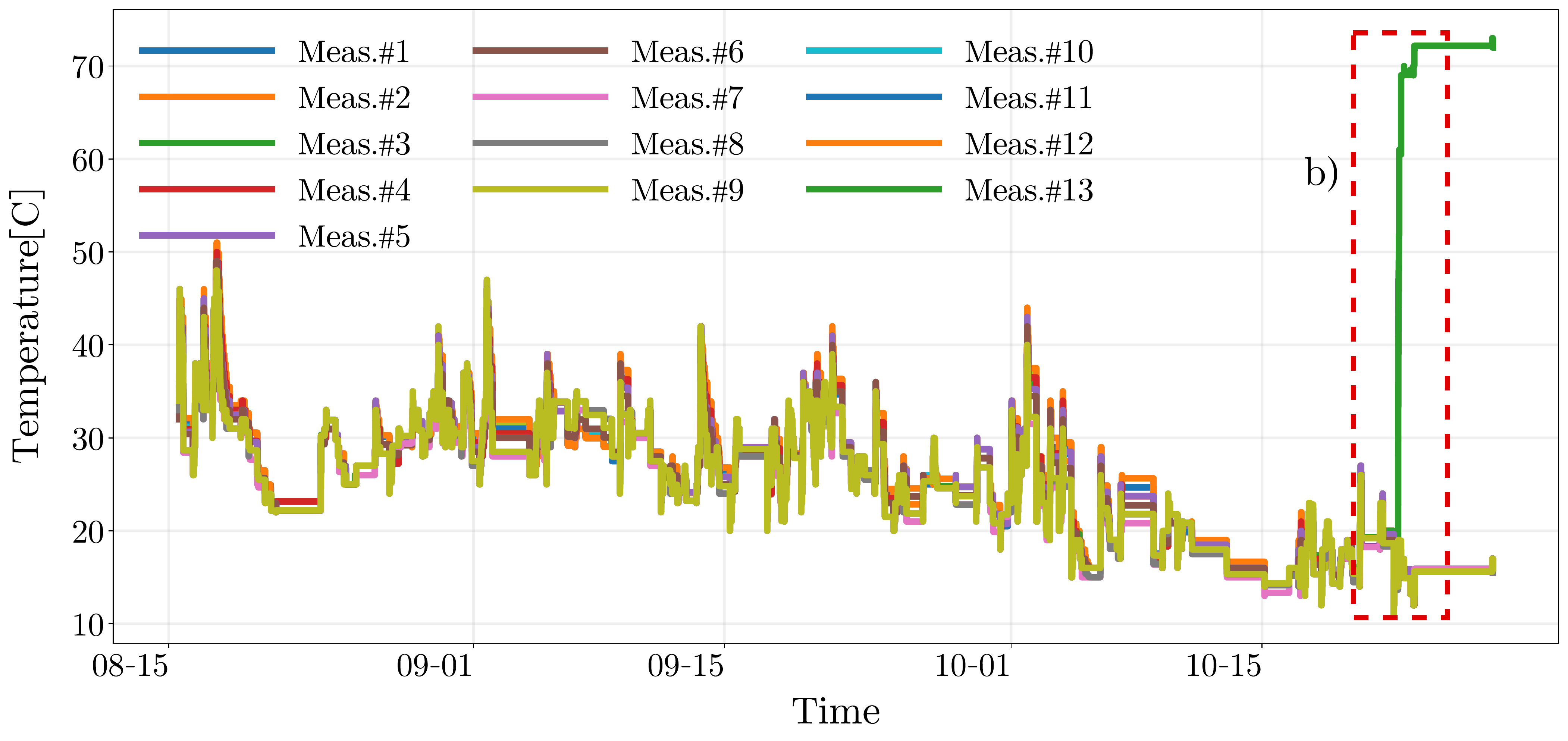}}
% leave a blank line to change row         
     \subfigure[anomaly detected 90min ahead]{\includegraphics[width=0.45\textwidth]{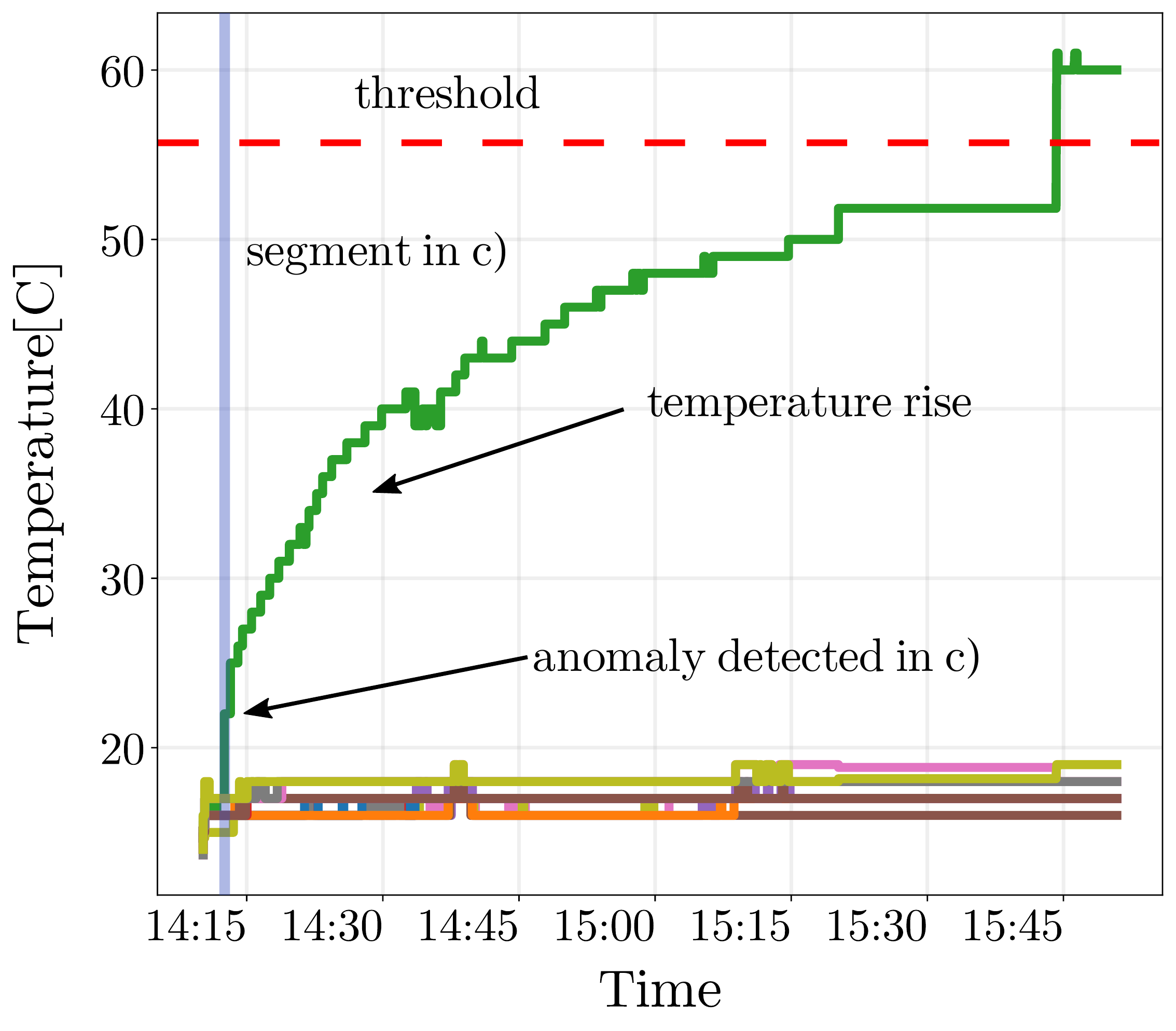}}
     \subfigure[one of the signal has different shape]{\includegraphics[width=0.45\textwidth]{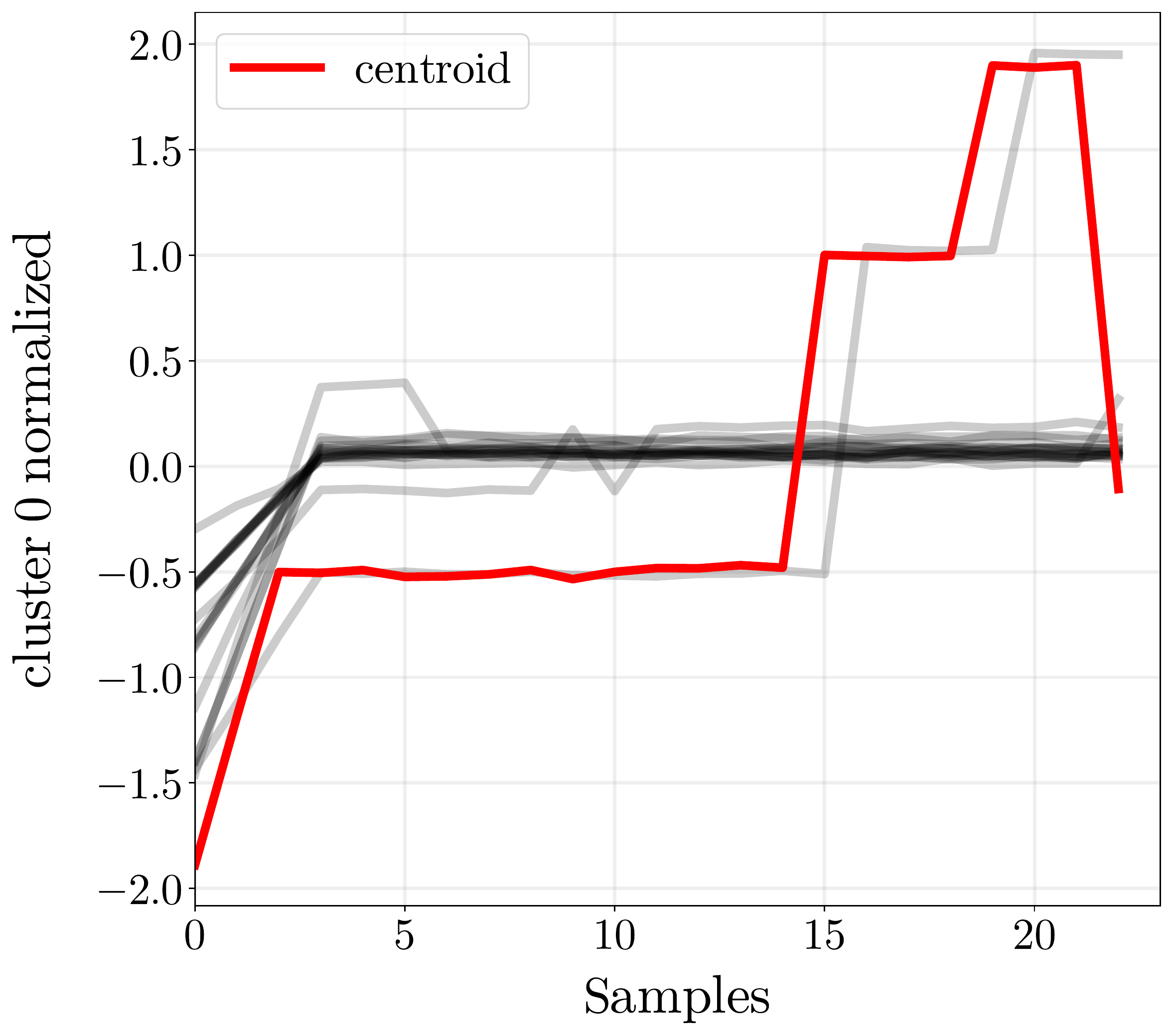}}
\caption{
Test Case (a) temperature measurement which shows the over-temperature fault. 
(b) The zoom-in view of the fault occurrence, this method detects the temperature anomaly 90min before the onboard BMS. (c) Further zoom-in view of the segment's shapes plot where the anomaly is detected, which is also highlighted with blue in (b)} \label{fig2}
\end{figure*}

We apply the proposed anomaly detection method to an EV battery. The data was collected and transmitted from an onboard BMS. As shown in Figure~\ref{fig2}(a), temperature measurement from sensor \#13 increased to over 70 \textcelsius on Oct 30th. The onboard BMS detects the over-temperature anomaly around 3:45 pm, during which the temperature was over 55 \textcelsius. Mean whole, the proposed method was able to detect anomalies around 2:15 pm, which is about 90 minutes earlier. The detailed comparisons are illustrated in Figure~\ref{fig2}(b). As it shows, the proposed method detect the anomaly when sensor \#13 just started to behave differently from the other measurements. Figure~\ref{fig2}(c) is the shape plot of the segment that detects the anomaly. In this figure, sensor \#13 is flagged as the outlier for its rising shape.

\section{Conclusions}

In this chapter, we present a Battery Cloud (Cloud Battery Management System) which is aimed at improving battery performance, safety, and economy by utilizing Cloud Computing and the Internet of Things. The major component of a battery cloud, include the  data sources during the stages of the battery life cycle, the different choices of databases, and deployment for battery data and data visualization. In addition, we discuss core algorithms for the Battery Cloud. Firstly, an artificial neural network is trained with cloud battery data for state-of-charge (SOC) estimation. The ANN is eventually deployed to the onboard BMS and tested on the vehicle. The  successful testing results show that cloud battery data is essential for developing advanced battery algorithms. Secondly, we discuss the degradation mechanisms of battery health and different algorithms for state-of-health (SOH). We develop a SOH estimation method based on DVA/ICA, which shows <5\% accuracy under different operating temperatures. At last, we review one important safety issue for batteries, thermal runaway. Its cause and effects are discussed. A data-driven battery anomaly detection method is developed to give early warnings. 

\bibliography{references}

% Generated by IEEEtran.bst, version: 1.14 (2015/08/26)
\begin{thebibliography}{10}
\providecommand{\url}[1]{#1}
\csname url@samestyle\endcsname
\providecommand{\newblock}{\relax}
\providecommand{\bibinfo}[2]{#2}
\providecommand{\BIBentrySTDinterwordspacing}{\spaceskip=0pt\relax}
\providecommand{\BIBentryALTinterwordstretchfactor}{4}
\providecommand{\BIBentryALTinterwordspacing}{\spaceskip=\fontdimen2\font plus
\BIBentryALTinterwordstretchfactor\fontdimen3\font minus
  \fontdimen4\font\relax}
\providecommand{\BIBforeignlanguage}[2]{{%
\expandafter\ifx\csname l@#1\endcsname\relax
\typeout{** WARNING: IEEEtran.bst: No hyphenation pattern has been}%
\typeout{** loaded for the language `#1'. Using the pattern for}%
\typeout{** the default language instead.}%
\else
\language=\csname l@#1\endcsname
\fi
#2}}
\providecommand{\BIBdecl}{\relax}
\BIBdecl

\bibitem{Dunn2011}
B.~Dunn, H.~Kamath, and J.-m. Tarascon, ``{Electrical energy storage for the
  Grid : A Battery of choices},'' \emph{Science Magazine}, vol. 334, no. 6058,
  pp. 928--936, 2011.

\bibitem{PacificNorthwestNationalLaboratory2022Lithium-ionNMC}
\BIBentryALTinterwordspacing
{Pacific Northwest National Laboratory}, ``{Lithium-ion Battery (LFP and
  NMC)},'' 2022. [Online]. Available:
  \url{https://www.pnnl.gov/lithium-ion-battery-lfp-and-nmc}
\BIBentrySTDinterwordspacing

\bibitem{Lombardo2021ArtificialReality}
T.~Lombardo, M.~Duquesnoy, H.~El-Bouysidy, F.~{\AA}r{\'{e}}n, A.~Gallo-Bueno,
  P.~B. J{\o}rgensen, A.~Bhowmik, A.~Demorti{\`{e}}re, E.~Ayerbe, F.~Alcaide,
  M.~Reynaud, J.~Carrasco, A.~Grimaud, C.~Zhang, T.~Vegge, P.~Johansson, and
  A.~A. Franco, ``{Artificial Intelligence Applied to Battery Research: Hype or
  Reality?}'' \emph{Chemical Reviews}, 2021.

\bibitem{Xu2014InternetSurvey}
L.~D. Xu, W.~He, and S.~Li, ``{Internet of things in industries: A survey},''
  \emph{IEEE Transactions on Industrial Informatics}, vol.~10, no.~4, 2014.

\bibitem{Voltaiq}
\BIBentryALTinterwordspacing
``{Voltaiq}.'' [Online]. Available: \url{https://www.voltaiq.com/}
\BIBentrySTDinterwordspacing

\bibitem{HowAWS}
\BIBentryALTinterwordspacing
``{How Gotion Monitors its EV Battery Solution with InfluxDB, Grafana and
  AWS}.'' [Online]. Available:
  \url{https://www.influxdata.com/resources/how-gotion-monitors-its-ev-battery-solution-with-influxdb-grafana-and-aws/}
\BIBentrySTDinterwordspacing

\bibitem{Schnell2019DataProduction}
\BIBentryALTinterwordspacing
J.~Schnell, C.~Nentwich, F.~Endres, A.~Kollenda, F.~Distel, T.~Knoche, and
  G.~Reinhart, ``{Data mining in lithium-ion battery cell production},''
  \emph{Journal of Power Sources}, vol. 413, no. October 2018, pp. 360--366,
  2019. [Online]. Available:
  \url{https://doi.org/10.1016/j.jpowsour.2018.12.062}
\BIBentrySTDinterwordspacing

\bibitem{ApacheHadoop}
\BIBentryALTinterwordspacing
{Apache}, ``{Apache Hadoop},'' 2022. [Online]. Available:
  \url{https://hadoop.apache.org/}
\BIBentrySTDinterwordspacing

\bibitem{AmazonService}
\BIBentryALTinterwordspacing
Amazon, ``{Amazon Web Service},'' 2022. [Online]. Available:
  \url{https://aws.amazon.com/}
\BIBentrySTDinterwordspacing

\bibitem{MicrosoftAzure}
\BIBentryALTinterwordspacing
Microsoft, ``{Microsoft Azure},'' 2022. [Online]. Available:
  \url{https://azure.microsoft.com/}
\BIBentrySTDinterwordspacing

\bibitem{GooglePlatform}
\BIBentryALTinterwordspacing
Google, ``{Google Cloud Platform},'' 2022. [Online]. Available:
  \url{https://cloud.google.com/}
\BIBentrySTDinterwordspacing

\bibitem{Cloudera}
\BIBentryALTinterwordspacing
``{Cloudera},'' 2022. [Online]. Available: \url{https://www.cloudera.com/}
\BIBentrySTDinterwordspacing

\bibitem{Influxdata}
\BIBentryALTinterwordspacing
influxdata, ``Influxdb,'' 2022. [Online]. Available:
  \url{https://www.influxdata.com/products/influxdb-overview/}
\BIBentrySTDinterwordspacing

\bibitem{Li2020DigitalEstimation}
\BIBentryALTinterwordspacing
W.~Li, M.~Rentemeister, J.~Badeda, D.~J{\"{o}}st, D.~Schulte, and D.~U. Sauer,
  ``{Digital twin for battery systems: Cloud battery management system with
  online state-of-charge and state-of-health estimation},'' \emph{Journal of
  Energy Storage}, vol.~30, no. April, p. 101557, 2020. [Online]. Available:
  \url{https://doi.org/10.1016/j.est.2020.101557}
\BIBentrySTDinterwordspacing

\bibitem{9699170}
A.~Abdollahi, J.~Li, X.~Li, T.~Jones, and A.~Habeebullah, ``{Voltage-Based
  State of Charge Correction at Charge-End},'' in \emph{2021 IEEE Vehicle Power
  and Propulsion Conference (VPPC)}, 2021, pp. 1--6.

\bibitem{Chemali2018State-of-chargeApproach}
\BIBentryALTinterwordspacing
E.~Chemali, P.~J. Kollmeyer, M.~Preindl, and A.~Emadi, ``{State-of-charge
  estimation of Li-ion batteries using deep neural networks: A machine learning
  approach},'' \emph{Journal of Power Sources}, vol. 400, no. June, pp.
  242--255, 2018. [Online]. Available:
  \url{https://doi.org/10.1016/j.jpowsour.2018.06.104}
\BIBentrySTDinterwordspacing

\bibitem{Hannan2017}
\BIBentryALTinterwordspacing
M.~A. Hannan, M.~S. Lipu, A.~Hussain, and A.~Mohamed, ``{A review of
  lithium-ion battery state of charge estimation and management system in
  electric vehicle applications: Challenges and recommendations},''
  \emph{Renewable and Sustainable Energy Reviews}, vol.~78, no. August 2016,
  pp. 834--854, 2017. [Online]. Available:
  \url{http://dx.doi.org/10.1016/j.rser.2017.05.001}
\BIBentrySTDinterwordspacing

\bibitem{Birkl.2017}
C.~R. Birkl, M.~R. Roberts, E.~McTurk, P.~G. Bruce, and D.~A. Howey,
  ``{Degradation diagnostics for lithium ion cells},'' \emph{Journal of Power
  Sources}, vol. 341, pp. 373--386, 2017.

\bibitem{Vetter.2005}
J.~Vetter, P.~Nov{\'{a}}k, M.~R. Wagner, C.~Veit, K.~C. M{\"{o}}ller, J.~O.
  Besenhard, M.~Winter, M.~Wohlfahrt-Mehrens, C.~Vogler, and A.~Hammouche,
  ``{Ageing mechanisms in lithium-ion batteries},'' \emph{Journal of Power
  Sources}, vol. 147, no. 1-2, pp. 269--281, 2005.

\bibitem{Dubarry.2012}
M.~Dubarry, C.~Truchot, and B.~Y. Liaw, ``{Synthesize battery degradation modes
  via a diagnostic and prognostic model},'' \emph{Journal of Power Sources},
  vol. 219, pp. 204--216, 2012.

\bibitem{Kabir2017DegradationReview}
M.~M. Kabir and D.~E. Demirocak, ``{Degradation mechanisms in Li-ion batteries:
  a state-of-the-art review},'' 2017.

\bibitem{Verma.2010}
\BIBentryALTinterwordspacing
P.~Verma, P.~Maire, and P.~Nov{\'{a}}k, ``{A review of the features and
  analyses of the solid electrolyte interphase in Li-ion batteries},''
  \emph{Electrochimica Acta}, vol.~55, no.~22, pp. 6332--6341, 2010. [Online].
  Available:
  \url{http://www.sciencedirect.com/science/article/pii/S0013468610007747}
\BIBentrySTDinterwordspacing

\bibitem{Aurbach.1999}
\BIBentryALTinterwordspacing
D.~Aurbach, B.~Markovsky, I.~Weissman, E.~Levi, and Y.~Ein-Eli, ``{On the
  correlation between surface chemistry and performance of graphite negative
  electrodes for Li ion batteries},'' \emph{Electrochimica Acta}, vol.~45,
  no.~1, pp. 67--86, 1999. [Online]. Available:
  \url{http://www.sciencedirect.com/science/article/pii/S0013468699001942}
\BIBentrySTDinterwordspacing

\bibitem{Keil.2017}
\BIBentryALTinterwordspacing
P.~Keil and A.~Jossen, ``{Aging of lithium-ion batteries in electric vehicles:
  Impact of regenerative braking},'' pp. 41--51, 2015. [Online]. Available:
  \url{https://mediatum.ub.tum.de/node?id=1355829}
\BIBentrySTDinterwordspacing

\bibitem{Broussely.2005}
\BIBentryALTinterwordspacing
M.~Broussely, P.~Biensan, F.~Bonhomme, P.~Blanchard, S.~Herreyre, K.~Nechev,
  and R.~J. Staniewicz, ``{Main aging mechanisms in Li ion batteries},''
  \emph{Journal of Power Sources}, vol. 146, no. 1-2, pp. 90--96, 2005.
  [Online]. Available:
  \url{http://www.sciencedirect.com/science/article/pii/S0378775305005082}
\BIBentrySTDinterwordspacing

\bibitem{PankajArora.1998}
\BIBentryALTinterwordspacing
P.~Arora, R.~E. White, and M.~Doyle, ``{Capacity Fade Mechanisms and Side
  Reactions in Lithium‐Ion Batteries},'' \emph{Journal of The Electrochemical
  Society}, vol. 145, no.~10, pp. 3647--3667, 1998. [Online]. Available:
  \url{https://iopscience.iop.org/article/10.1149/1.1838857}
\BIBentrySTDinterwordspacing

\bibitem{Han.2019}
X.~Han, L.~Lu, Y.~Zheng, X.~Feng, Z.~Li, J.~Li, and M.~Ouyang, ``{A review on
  the key issues of the lithium ion battery degradation among the whole life
  cycle},'' \emph{eTransportation}, vol.~1, p. 100005, 2019.

\bibitem{T.Waldmann.2014}
\BIBentryALTinterwordspacing
T.~Waldmann, S.~Gorse, T.~Samtleben, G.~Schneider, V.~Knoblauch, and
  M.~Wohlfahrt-Mehrens, ``{A Mechanical Aging Mechanism in Lithium-Ion
  Batteries},'' \emph{Journal of The Electrochemical Society}, vol. 161,
  no.~10, pp. A1742--A1747, 2014. [Online]. Available:
  \url{https://iopscience.iop.org/article/10.1149/2.1001410jes}
\BIBentrySTDinterwordspacing

\bibitem{Korthauer.2013}
R.~Korthauer, \emph{{Handbuch Lithium-Ionen-Batterien}}.\hskip 1em plus 0.5em
  minus 0.4em\relax Springer Berlin Heidelberg, 2013.

\bibitem{Hunt.1996}
\BIBentryALTinterwordspacing
{USABC}, ``{Electric Vehicle Battery Test Procedures - Rev. 2},'' 1996.
  [Online]. Available:
  \url{http://www.uscar.org/commands/files_download.php?files_id=73}
\BIBentrySTDinterwordspacing

\bibitem{Wood.2011}
\BIBentryALTinterwordspacing
E.~Wood, M.~Alexander, and T.~H. Bradley, ``{Investigation of battery
  end-of-life conditions for plug-in hybrid electric vehicles},'' \emph{Journal
  of Power Sources}, vol. 196, no.~11, pp. 5147--5154, 2011. [Online].
  Available:
  \url{http://www.sciencedirect.com/science/article/pii/S037877531100379X}
\BIBentrySTDinterwordspacing

\bibitem{V.Marano.2009}
V.~Marano, S.~Onori, Y.~Guezennec, G.~Rizzoni, and N.~Madella, ``{Lithium-ion
  batteries life estimation for plug-in hybrid electric vehicles},'' in
  \emph{5th IEEE Vehicle Power and Propulsion Conference, VPPC '09}, 2009, pp.
  536--543.

\bibitem{YanchengZhang.2011}
\BIBentryALTinterwordspacing
Y.~Zhang, C.~Y. Wang, and X.~Tang, ``{Cycling degradation of an automotive
  LiFePO4 lithium-ion battery},'' \emph{Journal of Power Sources}, vol. 196,
  no.~3, pp. 1513--1520, 2011. [Online]. Available:
  \url{https://pennstate.pure.elsevier.com/en/publications/cycling-degradation-of-an-automotive-lifeposub4sub-lithium-ion-ba}
\BIBentrySTDinterwordspacing

\bibitem{Tomaszewska.2019}
A.~Tomaszewska, Z.~Chu, X.~Feng, S.~O'Kane, X.~Liu, J.~Chen, C.~Ji, E.~Endler,
  R.~Li, L.~Liu, Y.~Li, S.~Zheng, S.~Vetterlein, M.~Gao, J.~Du, M.~Parkes,
  M.~Ouyang, M.~Marinescu, G.~Offer, and B.~Wu, ``{Lithium-ion battery fast
  charging: A review},'' \emph{eTransportation}, vol.~1, p. 100011, 2019.

\bibitem{Xiong.2018}
R.~Xiong, L.~Li, and J.~Tian, ``{Towards a smarter battery management system: A
  critical review on battery state of health monitoring methods},''
  \emph{Journal of Power Sources}, vol. 405, pp. 18--29, 2018.

\bibitem{Berecibar.2016b}
M.~Berecibar, I.~Gandiaga, I.~Villarreal, N.~Omar, J.~Van~Mierlo, and P.~Van
  Den~Bossche, ``{Critical review of state of health estimation methods of
  Li-ion batteries for real applications},'' \emph{Renewable and Sustainable
  Energy Reviews}, vol.~56, pp. 572--587, 2016.

\bibitem{Eddahech2014DeterminationPhase}
A.~Eddahech, O.~Briat, and J.~M. Vinassa, ``{Determination of lithium-ion
  battery state-of-health based on constant-voltage charge phase},''
  \emph{Journal of Power Sources}, vol. 258, 2014.

\bibitem{Dubarry2009IdentifyCell}
M.~Dubarry and B.~Y. Liaw, ``{Identify capacity fading mechanism in a
  commercial LiFePO4 cell},'' \emph{Journal of Power Sources}, vol. 194, no.~1,
  pp. 541--549, 10 2009.

\bibitem{Dubarry2011IdentifyingCells}
M.~Dubarry, B.~Y. Liaw, M.~S. Chen, S.~S. Chyan, K.~C. Han, W.~T. Sie, and
  S.~H. Wu, ``{Identifying battery aging mechanisms in large format Li ion
  cells},'' \emph{Journal of Power Sources}, vol. 196, no.~7, pp. 3420--3425, 4
  2011.

\bibitem{Han.2014}
\BIBentryALTinterwordspacing
X.~Han, M.~Ouyang, L.~Lu, and J.~Li, ``{A comparative study of commercial
  lithium ion battery cycle life in electric vehicle: Capacity loss
  estimation},'' \emph{Journal of Power Sources}, vol. 268, pp. 658--669, 2014.
  [Online]. Available:
  \url{http://www.sciencedirect.com/science/article/pii/S0378775314009756}
\BIBentrySTDinterwordspacing

\bibitem{Wang.2012}
Q.~Wang, P.~Ping, X.~Zhao, G.~Chu, J.~Sun, and C.~Chen, ``{Thermal runaway
  caused fire and explosion of lithium ion battery},'' \emph{Journal of Power
  Sources}, vol. 208, pp. 210--224, 2012.

\bibitem{Feng2018review}
X.~Feng, M.~Ouyang, X.~Liu, L.~Lu, Y.~Xia, and X.~He, ``{Thermal runaway
  mechanism of lithium ion battery for electric vehicles: A review},''
  \emph{Energy Storage Materials}, vol.~10, pp. 246--267, 1 2018.

\bibitem{Liao.2019}
Z.~Liao, S.~Zhang, K.~Li, G.~Zhang, and T.~G. Habetler, ``{A survey of methods
  for monitoring and detecting thermal runaway of lithium-ion batteries},''
  \emph{Journal of Power Sources}, vol. 436, p. 226879, 10 2019.

\bibitem{9576348}
X.~Li, J.~Li, A.~Abdollahi, T.~Jones, and A.~Habeebullah, ``{Data-driven
  Thermal Anomaly Detection for Batteries using Unsupervised Shape
  Clustering},'' in \emph{2021 IEEE 30th International Symposium on Industrial
  Electronics (ISIE)}, 2021, pp. 1--6.

\bibitem{Paparrizos2015}
J.~Paparrizos and L.~Gravano, ``{K-shape: Efficient and accurate clustering of
  time series},'' in \emph{Proceedings of the ACM SIGMOD International
  Conference on Management of Data}, vol. 2015-May.\hskip 1em plus 0.5em minus
  0.4em\relax Association for Computing Machinery, 5 2015, pp. 1855--1870.

\end{thebibliography}

\end{document}